\documentclass[preprint,secnumarabic,amssymb,amsmath,footinbib,tightenlines,apl,prb,showkeys,longbibliography]{revtex4-1}

\usepackage{amsmath,amssymb,amsfonts,textcomp,amsthm,mathtools}
\usepackage{gensymb}
\usepackage{setspace}  
\usepackage{marvosym}
\usepackage{psfrag}
\usepackage{subfigure}
\usepackage[dvips]{graphicx}
\usepackage{subfigure}
\usepackage{natbib}
\DeclareGraphicsExtensions{.eps}
\usepackage{color}
\usepackage{eulervm}
\usepackage{capt-of}
\usepackage{epstopdf}
\usepackage{array}
\usepackage{multirow}
\usepackage{ mathrsfs }
\usepackage{morefloats}
\usepackage{times}

\begin{document}

\title{Linear, nonlinear and transitional regimes of second mode instability}

\author{S. Unnikrishnan}
\affiliation{Mechanical Engineering, Florida State University, Tallahassee, Florida 32310} 

\author{Datta V. Gaitonde}
\affiliation{Mechanical and Aerospace Engineering, The Ohio State University, Columbus, Ohio 43210}

\date{\today}


\begin{abstract}
The 2D second-mode is a potent instability in hypersonic boundary layers (HBLs). 
We study its linear and nonlinear evolution, followed by its role in transition and eventual breakdown of the HBL into a fully turbulent state.
Linear stability theory (LST) is utilized to identify the second-mode wave through FS-synchronization, which is then recreated in linearly and nonlinearly forced 2D direct numerical simulations (DNSs). 
The nonlinear DNS shows saturation of the fundamental frequency, and the resulting superharmonics induce tightly braided ``rope-like'' patterns near the generalized inflection point (GIP). 
The instability exhibits a second region of growth constituted by the fundamental frequency, downstream of the primary envelope, which is absent in the linear scenario. 
Subsequent 3D DNS identifies this region to be crucial in amplifying oblique instabilities riding on the 2D second-mode ``rollers''. 
This results in lambda vortices below the GIP, which are detached from the ``rollers'' in the inner boundary layer. 
Streamwise vortex-stretching results in a localized peak in length-scales inside the HBL, eventually forming haripin vortices. 
Spectral analyses track the transformation of harmonic peaks into a turbulent spectrum, and appearance of oblique modes at the fundamental frequency, which suggests that fundamental resonance is the most dominant mechanism of transition. 
Bispectrum reveals coupled nonlinear interactions between the fundamental and its superharmonics leading to spectral broadening, and traces of subharmonic resonance as well. 
Global forms of the fundamental and subharmonic modes are extracted which shows that the former disintegrates at the location of spanwise breakdown, beyond which the latter amplifies.
Fundamental resonance results in complete breakdown of streamwise streaks in the lower log-layer, producing a fully turbulent HBL. Statistical analyses of near-wall flowfield indicate an increase in large-scale ``splatting'' motions immediately following transition, resulting in extreme skin friction events, which equilibrate as turbulence sets in.

\end{abstract}

\maketitle

\section{Introduction}\label{sec_intro}
Transition to turbulence is a major challenge in the aerodynamic design of vehicles. 
It causes boundary layers to induce higher wall-loading, resulting in drag and heat-transfer penalties. 
High-speed boundary layers exhibit several pathways to transition ({\em e.g.}, see \citet{morkovin1969many}), depending on the disturbance environment. 
Among these, the most explored are the linear mechanisms thought to be relevant in low-disturbance flight environment, which are driven by the instabilities of the base flow. 

It is well-known that the most aggressive instability of high-speed boundary layers (typically above Mach~$4$) is the so-called second-mode, or Mack-mode \citep{mack1975linear,mack1984boundary}. 
Being an inviscid instability of the base flow, this mode exhibits growth rates that are upto around $5$ times \citep{ozgen2008linear} of that observed in the Tollmien-Schlichting (T-S) (also referred to as the first-mode) instability, at Mach numbers relevant to hypersonic flight vehicles. 
Based on receptivity studies, the origin of second-mode in adiabatic hypersonic boundary layers (HBLs) can be traced back to a discrete mode, originating from the slow acoustic spectrum near the leading edge \citep{fedorov2003receptivity,fedorov2011transition}, with a phase-speed, $(1-1/M_\infty)$. 
The linear regime of this instability has also been addressed in several early experimental \citep{wilkinson1997review,schneider2013developing} and computational \citep{pruett1995spatial,ma2003receptivity1} studies. 

A through understanding and characterization of nonlinear and secondary-instability \citep{herbert1988secondary} regimes of the second-mode are also critical for reliable prediction and mitigation of transition in HBLs. 
Since the 2D waves of second-mode instability are the most highly amplified, turbulent breakdown is usually preceded by the development of oblique modes, as a result of secondary instability. 
Under natural conditions, the oblique modes generally gain prominence after the 2D waves are nonlinearly saturated. 
Such undulations of the second-mode could also be induced due to the presence of first mode instability, as seen in the direct numerical simulations (DNS) of \citet{khotyanovsky2016numerical}. 
Exploring this regime requires careful quantification of the mechanisms involved, and high-resolution data, to faithfully capture the relevant modal and non-modal interactions. 
\citet{craig2019nonlinear} explains the challenges associated with experimental quantification of this nonlinear regime, due to the limitations in frequency response of measurement systems, in the presence of superharmonics of the fundamental second-mode. 
The nonlinear evolution of second-mode instability within wavepackets in a Mach~$6$  boundary layer was studied using non-intrusive measurements by \citet{casper2014pressure}, which eventually broke down into turbulent spots. 
High-speed schlieren imaging has also enabled spectral, time-frequency and topology analyses on second-mode disturbances at hypersonic speeds, as reported in  \citet{laurence2016experimental}. 

Direct numerical simulations (DNSs), when anchored in experiments and interpreted along with constructs of linear theory, can provide crucial insights into the nonlinear and breakdown stages of second-mode instability. 
In addition, it provides a means to effectively control the initialization of specific candidate waves, to focus on predetermined routes to transition. 
For example, \citet{pruett1995spatial} studied the nonlinear stages of second-mode oblique waves using DNSs of HBLs over a Mach~$8$ cone. 
\citet{franko2013breakdown} simulated transition induced by second-mode instabilities through fundamental resonance and oblique mode breakdown, where an overshoot in heat transfer rates was observed for the latter route. 
Controlled excitation using wavepackets and harmonic waves by \citet{sv2014numerical,sv2015direct} have quantified the relative dominance of fundamental (K-type) resonance mechanisms in hypersonic cones over subharmonic (H-type) resonance.
Topological changes associated with enhanced compressibility of HBLs have also been identified as shown in \citet{jocksch2008growth}, where the near-wall region of nonlinear wavepackets contains spanwise coherent structures associated with second-mode instability, even in the late-transitional stages. 

In the current work, we perform a DNS-based study of the route to transition, following the evolution of a second-mode instability wave, in a Mach~$6$ boundary layer. 
Emphasis is on the key developments that transpire during the stages of linear amplification, nonlinear saturation, secondary instability and eventual breakdown into a fully turbulent state. 
A canonical flat-plate base flow is chosen to focus on the complexities in the modal dynamics of this instability. 
Due to the wide range of scales involved across the different stages, DNSs of these scenarios often adopt high-resolution numerics, with minimal artificial dissipation (see {\em e.g.}, \citet{ha2019dnsfc}). 
In \S\ref{sec_num}, we summarize the spatio-temporal schemes utilized for this study, along with the shock-resolution strategy. 
The controlled generation of second-mode instability waves using wall blowing-suction is informed by linear stability analysis (\S\ref{sec_lan}). 
Using the concept of FS-synchronization  \citep{ma2003receptivity1,fedorov2011high}, the streamwise location is identified where the phase-speeds of the slow and the fast discrete modes merge. 
This ensures that the instability growth downstream is induced due to the linear instability of the second-mode, constituted by the slow discrete mode (mode~S). 
\citet{wang2011response} identify the receptivity of HBLs which causes actuators that are located upstream of the FS-synchronization location to induce mode~S amplification at downstream locations, consistent with linear stability theory predictions.

DNSs can be effectively utilized to identify the multi-dimensional variations in the nonlinearly saturated second-mode instability, along with the distortions induced in the base flow. 
Modifications in the ``rope-shaped'' density-field structures of second-mode instability are reported in \citet{egorov2006direct} at nonlinear amplitudes, consistent with experimental observations. 
\S\ref{sec_nlev} addresses this regime using nonlinearly forced 2D DNS, and highlights major deviations from the linear response. 
The effects of saturation are further quantified by decomposing the nonlinear response into its orthogonal modes, representing the fundamental and its superharmonics. 
The numerical approach can also be seamlessly extended to the late transitional regime, to encompass realistic breakdown scenarios. 
This requires judicious choice of the 3D effects to be accounted for, to initiate the breakdown. 
Computations of secondary instabilities in a Mach~$4.5$ boundary layer by \citet{adams1996subharmonic} identify subharmonic resonance as a viable route to transition, when the 2D mode is allowed to naturally distort in the presence of random noise. 
Imposing specific spanwise wavenumebrs can realize controlled transition routes of fundamental resonance and oblique breakdown as shown in \citet{franko2013breakdown}. 
A simple random forcing approach by \citet{hader2018towards} to account for realistic wind-tunnel effects, resulted in transition over a flared cone, with spectral and heat-transfer characteristics consistent with controlled second-mode fundamental resonance. 
In this study, we adopt an approach in line with \citet{adams1996subharmonic}, where a 2D second-mode instability is excited in the HBL, in the presence of background random perturbations. 
This narrows-down the transition route to that initiated by second-mode instability, but allows the receptivity of the system to choose the preferred modes of secondary instabilities and eventual breakdown. 
The 3D simulation which captures the breakdown of the instability wave is presented in \S\ref{sec_bdtr}, where the vortical structures are analyzed to identify various stages of nonlinear evolution and length-scales present in the HBL.

Spectral characterization of the unsteady flowfield can help quantify crucial aspects like nonlinear saturation and breakdown. 
\citet{adams1996subharmonic} utilizes time-resolved and time-averaged frequency spectra to identity presence of second-mode wavepackets, as well as localized turbulent activity in HBLs. 
\S\ref{sec_bdtr} details spectral analyses, where the one-dimensional energy spectra is utilized to compare the DNS data to the turbulent spectrum. 
In addition, wavenumber-frequency analysis is performed to study the effect of oblique modes on the second-mode instability (see {\em e.g.} \citet{novikov2016direct}). 
While first-order spectra identify the presence of specific frequencies/wavenumbers, higher-order spectral analyses can yield insights into nonlinear coupled interactions that result in new frequencies/wavenumbers. 
\citet{kimmel1991nonlinear} have utilized a second-order spectral representation (bispectrum) to study nonlinear interactions within second-mode instabilities. 
We utilize the bispectrum to identify coupled interactions in the saturated second-mode, which results in harmonics, spectral broadening and mean flow distortions. 
Following this, the global form of the fundamental and subharmonic modes crucial to these interactions are extracted. 
The final section, \S\ref{sec_bdtr}, deals with the near-wall effects of transition, resulting in variations in skin-friction characteristics in the transitional and turbulent regions of the HBL. 
Correlation analyses and probability distributions are utilized to extract the dominant length-scales and qualitative trends in high-velocity patches, which form localized regions of intense skin friction.

\section{Numerics}\label{sec_num}
The governing equations are the unsteady 3D compressible Navier-Stokes equations, which are formulated for generalized curvilinear coordinates in the strong-conservation form:
\begin{equation} 
\frac{\partial}{\partial \tau} \biggl(\frac{\boldsymbol{Q}}{J}\biggl) = -\biggl[\biggl(\frac{\partial \boldsymbol{F_i}}
{\partial \xi} + \frac{\partial \boldsymbol{G_i}}{\partial \eta} + \frac{\partial \boldsymbol{H_i}}
{\partial \zeta}\biggl) + 
\frac{1}{Re} \biggl(\frac{\partial \boldsymbol{F_v}}{\partial \xi} + \frac{\partial \boldsymbol{G_v}}{\partial \eta} 
+ \frac{\partial \boldsymbol{H_v}}{\partial \zeta}\biggl) \biggl]                
\label{NSE}
\end{equation}      
The conserved variable vector is denoted by $\boldsymbol{Q}=[\rho,\rho u,\rho v, \rho w, \rho E]^T$. 
$\rho$ is density, $(u,v,w)$ are the Cartesian components of velocity, and $E={T}/{(\gamma-1){M_\infty}^2}+(u^2+v^2+w^2)/2$, is total specific internal energy. 
$T$ is temperature, ${M_\infty}$ is the reference freestream Mach number, and $\gamma$ is the ratio of the specific heats. 
The ideal gas law, $p=\rho T/{\gamma {M_\infty}^2}$, is assumed, where $p$ is pressure. 
$J=\partial{(\xi,\eta,\zeta,\tau)}/\partial{(x,y,z,t)}$ is the Jacobian of the coordinate transformation. 
Inviscid fluxes along the coordinates, ($\xi, \eta, \zeta$), are represented by, $(\boldsymbol{F_i}, \boldsymbol{G_i}, \boldsymbol{H_i})$. 
$(\boldsymbol{F_v}, \boldsymbol{G_v}, \boldsymbol{H_v})$ are the corresponding viscous fluxes. 
Further details on the formulation can be found in \citet{vinokur1974conservation,anderson2016computational}.

Reynolds number, $Re$, is defined as, $Re=\rho_\infty^{*}U_\infty^{*}L^{*}/\mu_\infty^{*}$, where $\mu_\infty^{*}$ is the freestream dynamic viscosity. 
Freestream conditions are utilized to normalize all variables except pressure, which is normalized as, $p=p^{*}/\rho_\infty^{*}{U_\infty^{*}}^2$.
$L^{*}$ is the reference length-scale. 
$(.)^*$ represents dimensional variables. 
Prandtl number, $Pr=0.72$, and $\gamma=1.4$. 
Sutherland's law is used to obtain temperature dependence of viscosity.

\begin{figure}
\centering
\setlength\fboxsep{0pt}
\setlength\fboxrule{0pt}
\fbox{\includegraphics[width=6.0in]{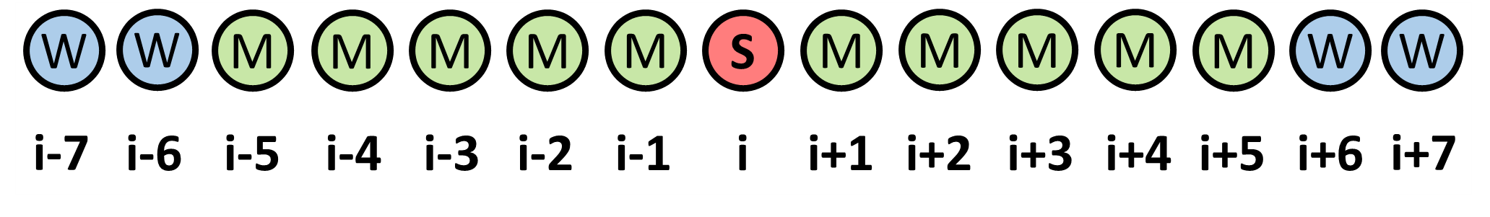}}
\caption{Schematic showing the shock-capturing technique. 
``S'' denotes the shock location. 
``M'' indicates shock vicinity, where third-order reconstruction is used. 
``W'' indicates locations where seventh-order reconstruction is used.}
\label{figskst}
\end{figure}

The discretized equations are solved in a finite difference framework using a high-order approach to ensure sufficient resolution of the wide range of scales involved in the transition phenomena. 
The high edge-Mach-number results in strong shocklets in the nonlinearly distorted and turbulent regions of the boundary layer, which calls-for robust shock-capturing schemes. 
These requirements are balanced using a shock detector routine\citep{bhagatwala2009modified}, which locally lowers the order of reconstruction in the vicinity of discontinuities. 
This is schematically represented in figure~\ref{figskst}.  
It represents a shock being detected at grid point, $i$, and is marked ``S''. 
Points from $i-5$ to $i+5$ (denoted ``M'') represent the vicinity of the shock, where primitive variables are reconstructed using a third-order MUSCL based scheme, along with the application of the van-Leer harmonic limiter \citep{lbv79-1}, to minimize grid-scale oscillations. 
At locations away from the shock (denoted ``W''), a seventh-order WENO reconstruction \citep{balsara2000monotonicity} is performed on the characteristic variables. 
The inviscid fluxes are then computed using the Roe scheme \citep{rpl81-1}. 
Viscous fluxes are discretized using the fourth-order central scheme. 
An implicit time-integration approach is adopted using the second-order diagonalized\citep{pth81-1} Beam-Warming approximate factorization\citep{br78-1}. 

\begin{figure}
\centering
\setlength\fboxsep{0pt}
\setlength\fboxrule{0pt}
\fbox{\includegraphics[width=6.0in]{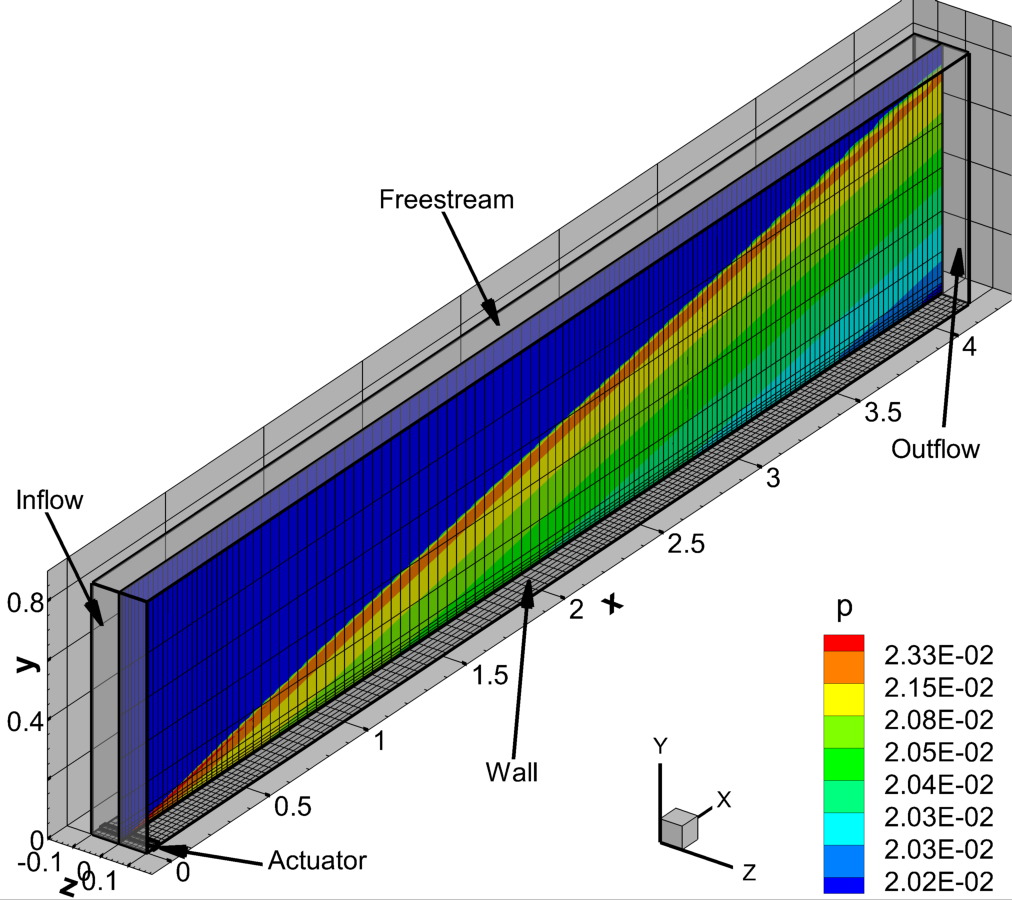}}
\caption{Computational domain along with the boundaries, as indicated. 
The location of the upstream actuator is also marked on the plate surface. 
A mid-span plane is shown along with pressure contours of the laminar flow.}
\label{figrd}
\end{figure}

The flowfield consists of a boundary layer, developing over an adiabatic flat plate with a sharp leading edge.   
The freestream conditions correspond to that described in \citet{egorov2006direct}, with $Re=2\times10^6$, and $M_\infty=6$. 
The computational domain and the laminar flowfield are presented in figure~\ref{figrd}, 
for reference. 
Every $20^{th}$ node is displayed in the wall-normal and spanwise directions, and every $40^{th}$ node is displayed in the streamwise direction, for clarity. 
$(x,y,z)$ are the Cartesian coordinates, corresponding to the streamwise, wall-normal and spanwise directions, respectively. 
The computational domain spans $0  \le x \le 4.2$, $0 \le y \le 0.85$ and $-0.1\le z \le 0.1$, where, $x=0$ coincides with the leading edge of the plate. 
Laminar pressure contours are also shown on the mid-span plane, and the wall-normal extent ensures that the leading edge shock is captured within the domain. 
The grid is clustered near the leading edge and the wall, and a uniform grid spacing is used in the spawise direction. 
A sponge zone is created through aggressive grid stretching beyond $x > 4$ and $y > 0.65$, to minimize reflections from the boundaries. 
The computational domain is resolved using $5531$, $301$ and $121$ nodes in the streamwise, wall-normal and spanwise directions, respectively. 
Based on fully turbulent boundary layer parameters at the outflow, the grid resolution in wall units are as follows: $\Delta x^+ = 2.5$, $\Delta y^+ = 0.3$ and $\Delta z^+ = 6.9$.  

The inflow plane is a supersonic inlet boundary, where freestream values are imposed on all primitive variables. 
Zero-streamwise-gradient condition is applied on the downstream outflow boundary.
Similarly, zero-wall-normal-gradient condition is applied on the freestream boundary. 
The surface of the plate is a no-slip, adiabatic wall. 
Periodic boundary condition is used in the spanwise direction. 
The actuator used to excite instabilities in the HBL is modeled as a blowing-suction slot, which introduces harmonic perturbations in wall-normal momentum, $q_w=\rho_w v_w$. 
Following, \citet{egorov2016direct}, this is defined as:
\begin{equation}
q_w(x,z,t)=\rho_w v_w=
A~sin\left(2\pi\frac{x-x_1}{x_2-x_1}\right)
sin(\omega_A t).
\end{equation}\label{eqn_wbsc}
The amplitude of the spanwise homogenous wave, $A$, depends on whether the analysis is linear or nonlinear, and is defined in relevant sections below. 
The frequency of forcing, $\omega_A$, and its upstream and downstream limits, $x_1$ and $x_2$, respectively, are obtained from linear stability analysis, as described in the following section.

The initial simulations used to characterize the linear and nonlinear properties of the second-mode instability are 2D in nature, which solves the 2D form of (\ref{NSE}). 
The spatial schemes, boundary conditions and the actuator model are identical to those described above for the 3D simulations. 
Time integration of the 2D equations are performed using the nonlinearly stable third-order Runge-Kutta scheme \cite{shu1988efficient}.

In the following sections, we study the evolution of the second-mode instability through the linear, nonlinear and transitional regimes.  
Hence, the first step is to identify the relevant wave parameters through linear analysis, that correspond to the second-mode instability in this HBL, as described below. 

\section{Linear analysis}\label{sec_lan}
Local linear stability theory (LST) is utilized to identify a wave that exhibits second-mode instability in this HBL. 
Due to its simplicity, a temporal framework is adopted to obtain an estimate of the unstable frequencies and corresponding wavelengths \citep{malik1990numerical}. 
Briefly, the Navier-Stokes equations are linearized following the Reynolds decomposition, and the laminar basic state is assumed to be one dimensional. 
The perturbations are composed of waves that are defined by streamwise and spanwise wavenumbers, $\alpha$ and $\beta$, respectively, with circular frequency, $\omega$. 
The Reynolds decomposition and the wave ansatz for any primitive variable, $\phi$, can be represented as:
\begin{equation}\label{pois_psia1f_1}
\phi = \bar{\phi} + \phi ', \quad
\phi ' = \hat{\phi}(y)e^{i(\alpha x + \beta z - \omega t)}.
\end{equation}
$\bar{(.)}$ and $(.)'$ are time-averaged and perturbation quantities, respectively. 
The resulting eigenvalue problem is solved to obtain the complex eigenvalues, $\omega$, for specified real wavenumbers, $\alpha$ and $\beta$. 
Since two-dimensional waves exhibit the highest growth rates in second-mode instability \citep{malik1990numerical,yao2007effect}, we use $\beta=0$. 

\begin{figure}
\centering
\setlength\fboxsep{0pt}
\setlength\fboxrule{0pt}
\fbox{\includegraphics[width=6.0in]{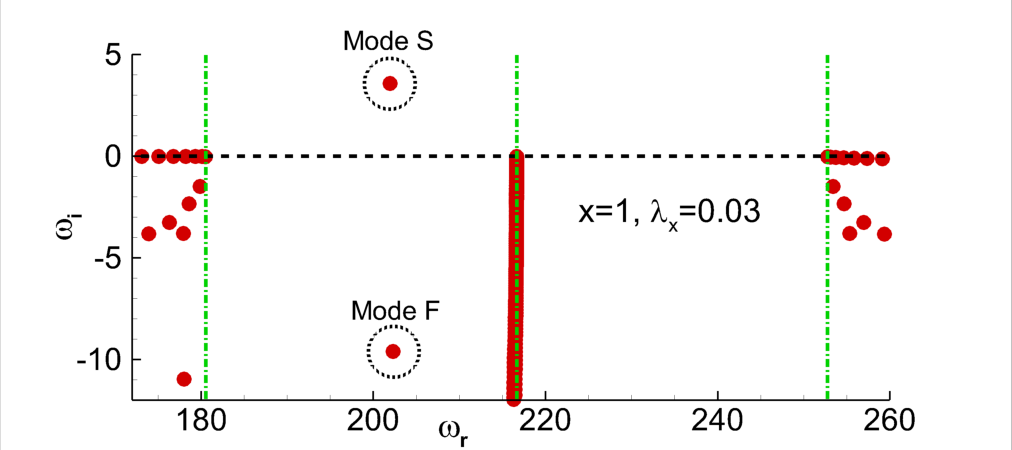}}
\caption{Eigenspectrum from linear stability analysis obtained at $x=1$. 
The left and right vertical lines correspond to the slow and the fast acoustic speed limits, respectively. 
The unstable mode~S and the stable mode~F are marked with circles, in the vicinity of their phase-speed synchronization.}
\label{figeigspc}
\end{figure}

For adiabatic walls, the second-mode instability is exhibited by the so called mode~S, which is a discreet eigenmode originating from the slow continuous acoustic spectrum (see {\em e.g.} \citet{fedorov2003receptivity,fedorov2011transition}). 
With increasing downstream distance from the leading edge, as the growth rate of this mode becomes positive (unstable), the phase-speed of mode~S synchronizes with that of mode~F, which is another discreet mode branching-off from the fast continuous acoustic spectrum \citep{ma2003receptivity1,fedorov2011high}. 
This phenomenon helps identify the unstable second-mode instability, and is shown in figure~\ref{figeigspc}, 
which is the eigenspectrum of the basic state at $x=1$. 
The horizontal and vertical axes are the real (circular frequency) and imaginary (growth rate) parts of the eigenvalue, $\omega_r$ and $\omega_i$, respectively. 
The left, middle and right vertical dash-dot lines correspond to the continuous slow-acoustic, vortical/entropic and fast-acoustic branches, respectively. 
The two circled eigenvalues represent mode~S and mode~F that have synchronized, as indicated by their abscissa. 
As expected, mode~S has a positive growth rate, and is the second-mode instability of interest here. 
The eigenspectrum is obtained for a perturbation wave with a streamwise wavelength, 
$\lambda_x = 2 \pi / \alpha = 0.03$.

(a) Variation of phase speeds, $c_r$, of mode~F and mode~S with streamwise distance from the leading edge. 
The top, middle and bottom dashed black lines mark the phase speeds of the fast-acoustic, vortical/entropic and slow-acoustic waves, respectively. 
(b) Variation of amplification rates, $c_i$, of mode~F and mode~S with streamwise distance from the leading edge.

\begin{figure}
\centering
\setlength\fboxsep{0pt}
\setlength\fboxrule{0pt}
\fbox{\includegraphics[width=6.0in]{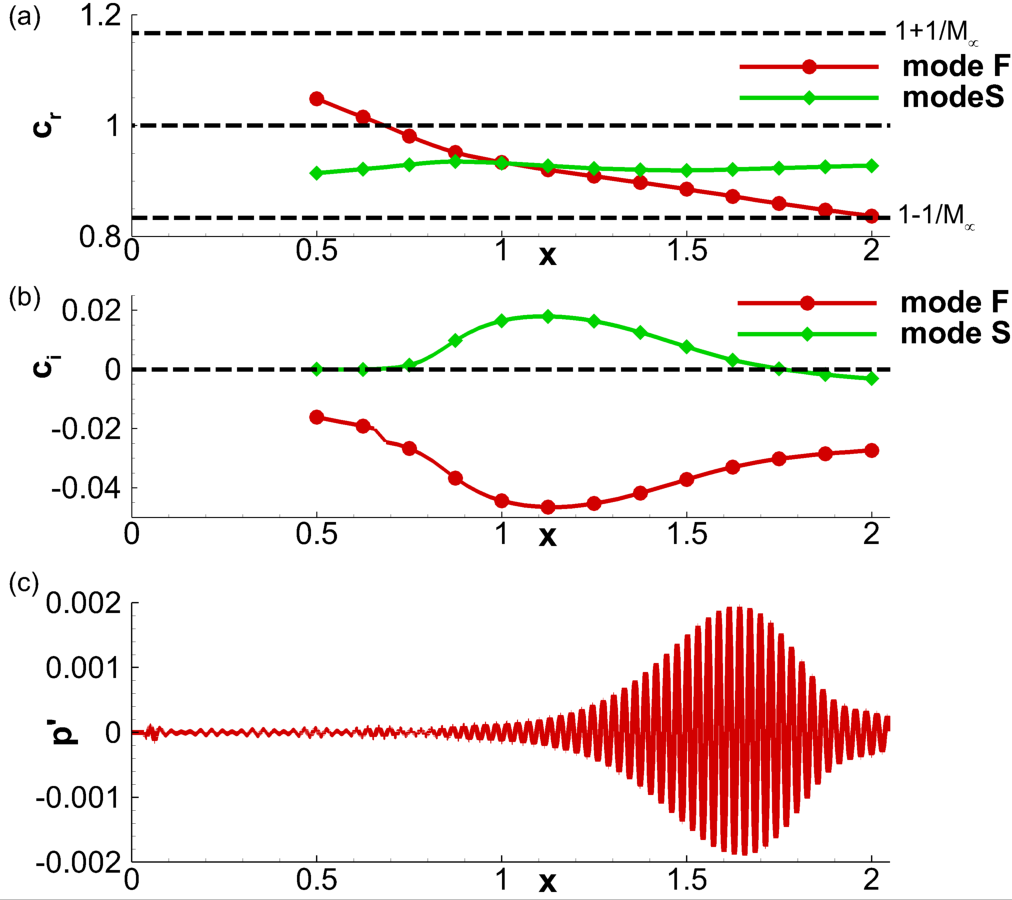}}
\caption{(a) Variations in phase-speed, $c_r$, of mode~F and mode~S  as a function of streamwise distance from the leading edge. 
(b) Variations in growth rate, $c_i$, of mode~F and mode~S  as a function of streamwise distance from the leading edge. 
(c) Linear response of DNS to second-mode actuation shown using wall-pressure perturbation.}
\label{figeigsfs}
\end{figure}

The above mentioned FS-synchronization is evident in the phase-speed plots of mode~S and mode~F, presented in figure~\ref{figeigsfs}. 
Figure~\ref{figeigsfs}(a) plots the loci of (real) phase-speed, $c_r= \omega_r/\lambda_x$, as a function of streamwise distance from the leading edge. 
The phase-speed of mode~S is found to be relatively invariant beyond $x=0.5$, whereas that of mode~F gradually decreases from the fast acoustic limit, $(1+1/M_\infty)$, to the slow acoustic limit, $(1-1/M_\infty)$, as observed in \citet{ma2003receptivity1}. 
During this process, mode~F first intersects with the continuous vortical/entropy spectrum \citep{fedorov2003initial} at $c_r=1$, and then, mode~S. 
The loci of the imaginary parts of phase-speeds are plotted in figure~\ref{figeigsfs}(b). 
In the vicinity of FS-synchronization, the growth rate of mode~S becomes positive ($c_i > 0$), and eventually reaches peak values around $x \sim 1.1$. 
For the freestream conditions under consideration, the unstable growth rates of mode~S after FS~synchronization can reach magnitudes five to ten times (see {\em e.g.} \citet{ozgen2008linear,unnikrishnan2019first}) higher than that prior to it. 
Mode~F remains damped over this adiabatic wall, since $c_i < 0$. 

The eventual nonlinear evolution of second-mode will be studied through DNS. 
Therefore, it is important to first quantify the linear response of DNS, to reconcile the differences with LST, induced due to the temporal framework and 1D assumption of mean flow in the latter approach. 
In addition, the DNS basic state is also slightly altered due to viscous-inviscid interactions near the leading edge of the plate. 
For this, a 2D DNS is performed on the plane $z=0$, to obtain a converged laminar flow (previously shown in figure~\ref{figrd}). 
This flow is then perturbed by small-amplitude wall-normal blowing-suction as defined by (\ref{eqn_wbsc}), with $A=5\times10^{-4}$. 
Similar amplitudes have been used by \citet{egorov2006direct} to obtain linear second-mode response in 2D DNS. 
It is seen in figure~\ref{figeigspc} that, the circular frequency of mode~S is around $\omega_r = 200$ in the vicinity of FS~synchronization. 
Hence, the actuator frequency is also chosen as $\omega_A=200$. 
The streamwise extent of the actuator slot is defined as $x_1=0.035$, and $x_2=0.064$, which is approximately equal to the wavelength of the instability wave identified in figure~\ref{figeigspc}. 
As observed in \citet{zhong2001leading, wang2009effect}, when the actuator is placed upstream of the FS~synchronization point, the unstable mode~S is naturally excited in the HBL.

The linear response of second-mode instability thus obtained in the DNS is reported in figure~\ref{figeigsfs}(c) to facilitate a direct comparison with corresponding LST results. 
Wall-pressure perturbation, $p'$, is plotted with $x$, in the region of FS~synchronization. 
The most significant amplification in linear DNS coincides with the unstable region of mode~S, with peak amplitude observed at $x \sim 1.65$. 
This is also consistent with the streamwise location at which mode~S amplification rate falls below $c_i =0$, thus indicating the location at which second-mode instability begins to attenuate. 
The reasonable agreement between LST and linear DNS ensures that the wave parameters chosen, induce a second-mode instability in the numerical simulations,  within the computational domain.

\begin{figure}
\centering
\setlength\fboxsep{0pt}
\setlength\fboxrule{0pt}
\fbox{\includegraphics[width=6.0in]{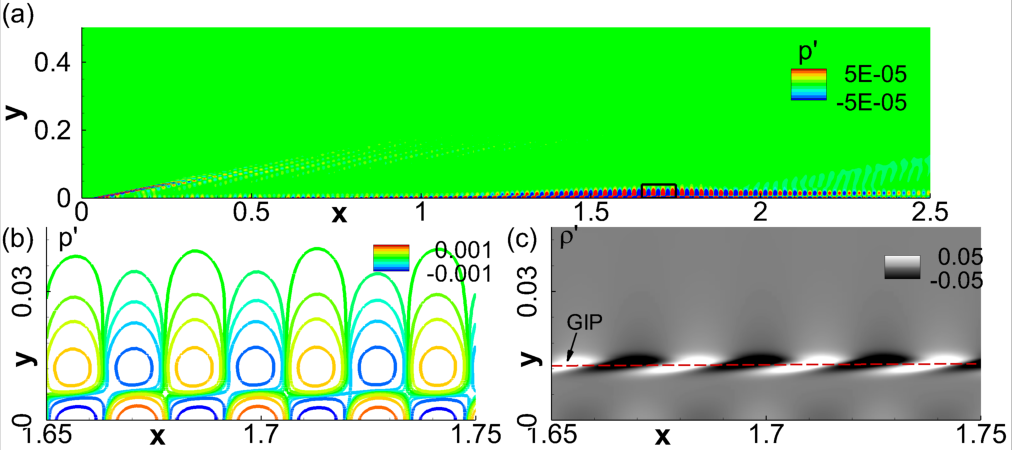}}
\caption{(a) Pressure perturbation contours in the linear DNS. 
(b) Magnified view of pressure perturbation contours in the region of second-mode amplification (marked in (a) with a rectangle). 
(c) Corresponding contours of density perturbation field. 
The dashed line in (c) represents the GIP.}
\label{figlndns}
\end{figure}

The characteristic features of second-mode instability are presented in figure~\ref{figlndns}. 
The pressure perturbation contours in figure~\ref{figlndns}(a) shows that instability amplification is bounded within the boundary layer, primarily in the zone $1.2<x<2.2$. 
The signature of the actuator is also visible in the vicinity of the leading edge. 
The pressure perturbation contours in the vicinity of peak amplification (marked by a rectangle in figure~\ref{figlndns}(a)) are plotted in detail in figure~\ref{figlndns}(b). 
The classic two-lobbed structure of second-mode instability waves is evident with compact wall-normal support. 
The density perturbations visualized in figure~\ref{figlndns}(c) exhibit ``rope-shaped'' patterns, observed commonly in experimental measurements \citep{stetson1993breakdown,laurence2016experimental,kennedy2018visualization} . 
These peak levels of density perturbations align with the generalized inflection point (GIP) in the mean profile, where the gradients are generally the highest, and engender aggressive perturbation growth. 
The locus of GIP is also marked using a horizontal dashed line. 
The outer lobes in pressure perturbations also occur at this wall-normal location, as can be seen by comparing figures~\ref{figlndns}(b) and(c). 

\section{Nonlinear evolution of second-mode}\label{sec_nlev}
\begin{figure}
\centering
\setlength\fboxsep{0pt}
\setlength\fboxrule{0pt}
\fbox{\includegraphics[width=6.0in]{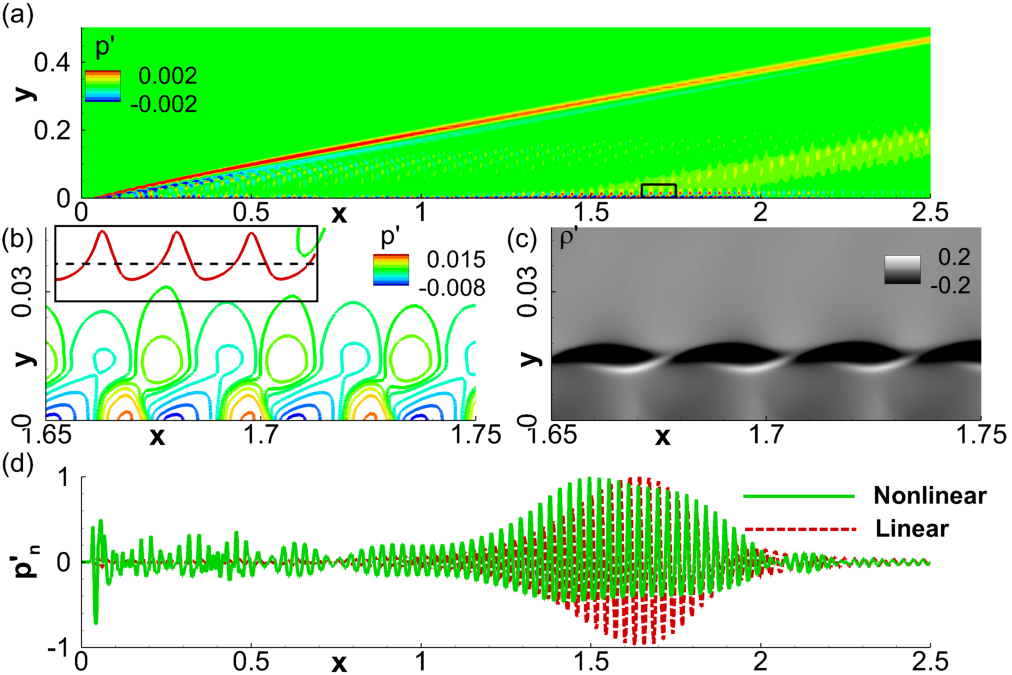}}
\caption{(a) Pressure perturbation contours in the nonlinear DNS. 
(b) Magnified view of pressure perturbation contours in the region of second-mode amplification (marked in (a) with a rectangle). 
Inset in (b) plots surface pressure perturbations with respect to the zero-mark (dashed horizontal line). 
(c) Corresponding contours of density perturbation field. 
(b) Normalized surface pressure perturbations in the linear and nonlinear DNSs.}
\label{fignnlndns}
\end{figure}

Prior to analyzing the breakdown scenario, it is illustrative to look at the saturated second-mode, so that the effects of nonlinearity on the 2D wave are evident. 
For this, 2D forced DNS is performed with an actuator amplitude, $A=5\times10^{-2}$. 
The resulting perturbation field is reported in figure~\ref{fignnlndns}.  
The pressure perturbation contours in figure~\ref{fignnlndns}(a) indicate waves accumulating behind the shock wave, as well as amplifying within the boundary layer.  
Unlike the linear scenario, pressure perturbations here show occasional extensions into the freestream, as was also seen in the nonlinear behavior of second-mode by \citet{egorov2016direct}.
The region marked by the rectangle (in figure~\ref{fignnlndns}(a)) is magnified in figure~\ref{fignnlndns}(b) for a detailed representation of the pressure perturbation contours. 
The pressure cells are distorted in an alternating pattern, with the corresponding waveform exhibiting wider troughs and narrower steep peaks. 
The surface pressure is plotted in the inset of figure~\ref{fignnlndns}(b) in the range, $1.65 \le x \le 1.75$, for a qualitative representation of this distortion. 
The horizontal dashed line is the zero-mark. 
The density perturbation contours in this region are presented in figure~\ref{fignnlndns}(c). 
While the general ``rope-shaped'' patterns persist, there are some variations from the linear structures observed in figure~\ref{figlndns}(c). 
For example, the perturbations in the linear field are symmetric about the zero value, as indicated by the identical shapes of the black and white zones. 
The nonlinear field is asymmetric, with the negative deviations being spatially dominant. 
The interlacing is also pronounced in the nonlinear response \citep{egorov2016direct}, with the braided pattern becoming more evident. 
This is seen for instance, in the experiments of \citet{laurence2016experimental}, where the initial stages of second-mode wavepackets display symmetric fluctuations. 
The tightly braided structures become evident upon tracking these wavepackets into the nonlinearly saturated regime. 

Nonlinearty can alter the growth-envelope of second-mode instabilities from that predicted by LST. 
To quantify this, we plot the surface pressure perturbations from the linear and nonlinear DNSs in figure~\ref{fignnlndns}(d). 
These pressure plots are normalized by their respective peak absolute values to obtain $p'_n$, to facilitate a straightforward comparison. 
Due to the high growth rate of the second-mode instability in the linear regime, the  primary zone of amplification between $1.2<x<2.2$ essentially masks the pressure trace elsewere over the wall. 
In contrast, nonlinear saturation limits peak amplification to around $3$ to $5$ times of that observed at upstream locations. 
In the nonlinear case, second-mode achieves peak amplitudes at an upstream location, compared to the linear case. 
This could be due to the energization of superharmonics with smaller wavelengths, that can be harbored in a thinner boundary layer upstream. 
The envelope of the nonlinear wave is also asymmetric and modulated as can be seen in the range, $1.5 \le x \le 2$. 
Such modulations are characteristic of nonlinear saturation \citep{ha2020threed} and will be further examined in the spectral domain, below. 
Beyond $x=2$ the linear response monotonically decays, whereas the nonlinear response exhibits a second region of amplification between $2 \le x \le 2.5$. 
These regions become relevant in the 3D simulation discussed in the next section, by influencing oblique mode instabilities and facilitating the breakdown of the HBL.
 
\begin{figure}
\centering
\setlength\fboxsep{0pt}
\setlength\fboxrule{0pt}
\fbox{\includegraphics[width=6.0in]{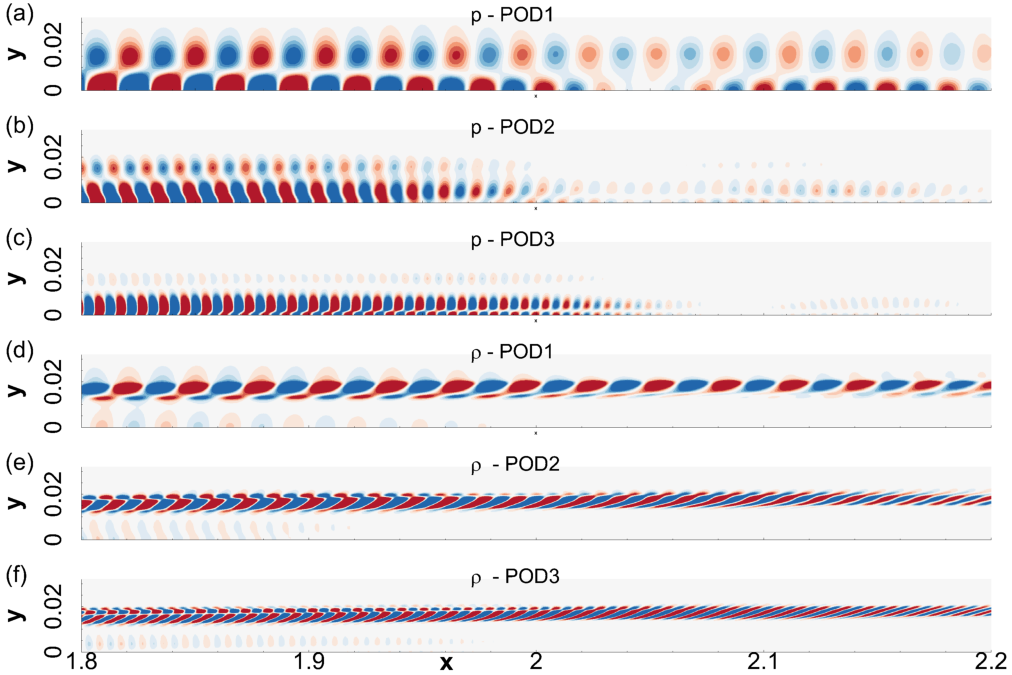}}
\caption{(a) First, (b) second and (c) third orthogonal modes in pressure perturbations. 
(d) First, (e) second and (f) third orthogonal modes in density perturbations.}
\label{fignlpod}
\end{figure}

The differences between the linear and nonlinear perturbation fields are manifested as superharmonics of the primary (forcing) frequency. 
To better quantify these differences, we split the pressure and density perturbations fields into orthogonal modes using proper orthogonal decomposition (POD). 
Since the nonlinear field was ensured to be composed of the forcing frequency and its integer superharmonics alone, the POD modes naturally coincide with these harmonics. 
This was verified post-factum by obtaining the frequency spectra of the monochromatic POD modes. 
Due to the harmonic nature of the perturbations, the POD modes appeared in pairs, the first three of which are presented in figure~\ref{fignlpod}, 
for pressure (p-POD1, p-POD2 and p-POD3) and density ($\rho$-POD1, $\rho$-POD2 and $\rho$-POD3). 
They correspond to modes at frequencies, $\omega_A$, $2\omega_A$ and $3\omega_A$, in the respective primitive variables. 
The streamwise extent is chosen so as to highlight the deviation from the linear response, in $1.8 \le x \le 2.2$. 
The dual-lobed pressure contours and the ``rope-shaped'' density patterns in the leading modes of the nonlinear field recover the corresponding linear response observed earlier in figure~\ref{figlndns}. 
In addition to this, the primary mode in pressure most clearly accounts for the second region of amplification observed in the nonlinear response in $2 \le x \le 2.5$. 
The higher modes of pressure also exhibit a dual-lobed structure in the region of peak amplitudes of the primary mode, and are increasingly restricted to within the boundary layer. 
The first and second superharmonics in density perturbations are also confined to the vicinity of the GIP, and exhibits a higher degree of interlacing, resulting in the braided features in the overall nonlinear response (figure~\ref{fignnlndns}(c)). 
In all the primitive variables examined, the wavelengths of higher modes decrease by the same factor at which the frequencies increase in the superharmonics, thus imparting similar phase-speeds to all the modes. 

Following the analysis of linear and nonlinear behavior of the 2D second-mode instability, we now perform 3D DNS to identify mechanisms leading to its breakdown and eventual transition of the HBL.

\section{Breakdown and transition}\label{sec_bdtr}
For simulating transition, the amplitude of the spanwise homogeneous wave in the 3D DNS is maintained the same ($A=5\times10^{-2}$) as in the nonlinear 2D DNS analyzed above. 
To capture the receptivity of the nonlinearly distorted HBL to oblique instabilities, a background random perturbation field is also imposed on the actuator. 
It also enhances the stochastic nature of unsteadiness post breakdown in the HBL \citep{sayadi2013direct}. 
This random field is varied across three orders of magnitude to ensure that the breakdown characteristics are not sensitive to its amplitude. 
In the reported results, the root mean square value of random perturbations are of the order, $1\times10^{-4}$. 

\begin{figure}
\centering
\setlength\fboxsep{0pt}
\setlength\fboxrule{0pt}
\fbox{\includegraphics[width=6.0in]{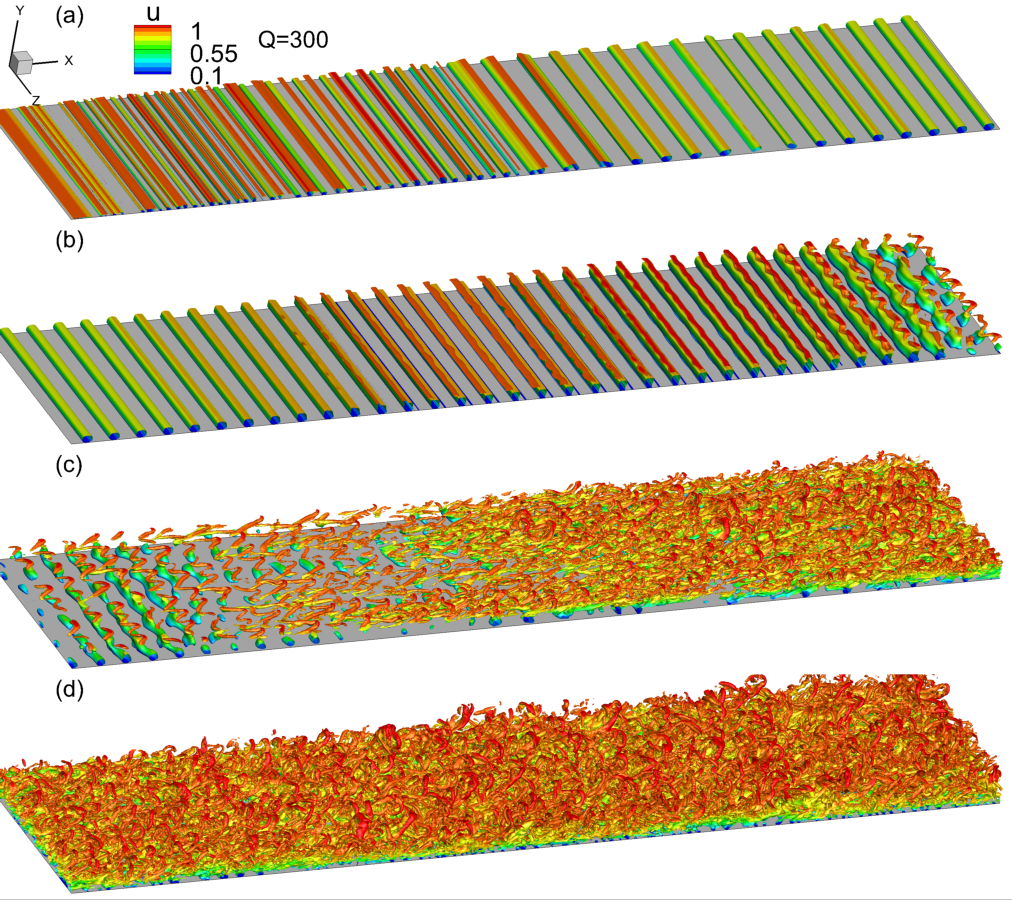}}
\caption{DNS results of transition visualized using Q-criterion, colored with $u$. 
Each frame represents equal streamwise segments, as follows: (a) $0 \le x \le 1$, (b) $1 \le x \le 2$, (c) $2 \le x \le 3$ and (d) $3 \le x \le 4$.}
\label{figqiso}
\end{figure}

Overall features of the destabilized HBL are presented first, in figure~\ref{figqiso}, 
using an iso-level of Q-criterion, colored by $u$. 
Each frame represents equal streamwise segments of the plate defined as, $0 \le x \le 1$, $1 \le x \le 2$, $2 \le x \le 3$ and $3 \le x \le 4$, for figures~\ref{figqiso}(a), (b), (c) and (d), respectively. 
The following observations are made from the vortical structures highlighted through Q-criterion. 
Further quantitative evidences are provided in the context of relevant discussions in the following sections. 
\begin{itemize}

\item In the range, $0 \le x \le 1$, the behavior of the second-mode instability in the 3D simulation is essentially same as to that observed in the 2D nonlinear simulation, discussed above. 
The HBL is dominated by spanwise homogeneous ``rollers'' which forms distortions in the laminar basic state over which secondary instabilities evolve. 

\item Although the ``rollers'' still remain the dominant coherent feature in $1 \le x \le 2$, traces of spanwise variations begin to appear near the GIP. 
This is evident by $x \sim 1.5$, prominently on the freestream-side of the ``rollers'', compared to the near-wall region. 
By  $x \sim 2$, the top region of the ``rollers'' is sufficiently modulated with the most receptive oblique waves to detach from the 2D waves in the inner boundary layer, evolving into lambda vortices. 
In the near-wall region, the ``rollers'' are eventually distorted (to a lesser extent) and initiates spanwise breakdown. 

\item As seen in $2 \le x \le 3$ the spanwise breakdown rapidly destabilizes the HBL, forming hairpin vortices, characteristic of initial stages of turbulence in boundary layers \citep{schlatter2010assessment}. 
Detailed analysis of the coherent structures in $2 \le x \le 2.5$ indicates that the ``legs'' of lambda vortices consisting of streamvise vortex tubes stretches in the streamwise direction. 
This is accompanied by the ``inclination'' \citep{jeong1997coherent} of these vortices such that the ``head'' region moves away from the wall. 
Towards $x \sim 3$, we observe densely arranged hairpin vortices in the boundary layer, and a rapid broadening of spanwise length-scales in the flow. 
Wall-shear measurements indicated that the average skin friction increases by a factor of around $4$ within the range, $2.5 \le x \le 3$.
Due to the random nature of spanwise inhomogenity seeded with the actuator, complete breakdown of the boundary layer occurs at slightly different streamwise locations across the span in an instantaneous sense.

\item The final section, $3 \le x \le 4$, displays a turbulent boundary layer, composed of broadband content in the frequency and waveumber domains. 
The ``forest of hairpins'' \citep{wu2009direct} in early-turbulence ($x \sim 3$) indicates that the boundary layer has attained statistical invariance in the spanwise direction. 
Further downstream, the boundary layer thickens and ``arch-like'' structures appear in the outer layer \citep{eitel2015hairpin} , which are detached from the streamwise vortices observed closer to the wall \citep{jeong1997coherent}. 
In addition, beneath the streamwise vortices, near-wall spanwise-oriented structures also appear, which have been previously observed in transitional regions \citep{jocksch2008growth} above $M_\infty=5$, and are found here to persist into the fully-developed turbulent region of the HBL.

\end{itemize} 

\begin{figure}
\centering
\setlength\fboxsep{0pt}
\setlength\fboxrule{0pt}
\fbox{\includegraphics[width=6.0in]{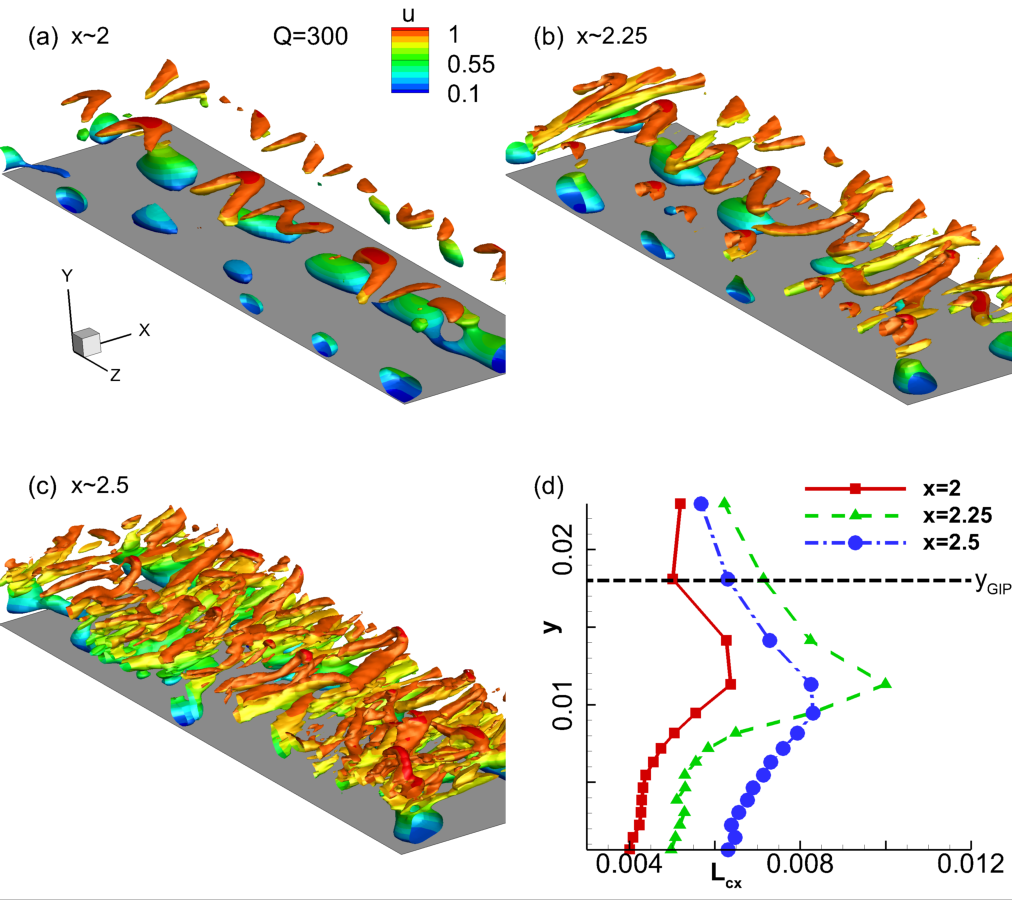}}
\caption{Q-criterion colored with $u$, highlighting behavior of oblique instabilities in the region of transition: (a) $x \sim 2$, (b) $x \sim 2.25$, (c) $x \sim 2.5$. 
(d) Variation of streamwise length-scales inside the boundary layer at the indicated locations.}
\label{figqiso3lc}
\end{figure}

Since initiation of breakdown from the 2D wave is of direct interest to this study, further details of spanwise inhomogenity in the ``rollers'' are presented in figure~\ref{figqiso3lc}. 
The upstream region of figure~\ref{figqiso}(c), $2 \le x \le 2.5$, is highlighted through the iso-level of Q-criterion, colored by $u$. 
Three locations are chosen, in the vicinity of $x \sim 2$, $x \sim 2.25$ and $x \sim 2.5$, to characterize the early spanwise variations in the second-mode. 
The formation of lambda vortices near the GIP is evident in figure~\ref{figqiso3lc}(a), with the ``rollers'' beneath it. 
Their ``inclination'' relative to the wall increases downstream, as seen in figure~\ref{figqiso3lc}(b). 
The stretching of the streamwise vortex filaments results in hairpin vortices with the ``head'' region in the outer boundary layer (figure~\ref{figqiso3lc}(c)).

The ``inclination'' and stretching of vortices can be verified from the trends in the streamwise length-scales present across the height of the boundary layer. 
This is quantified in figure~\ref{figqiso3lc}(d) using the streamwise integral length-scale, $L_{cx}$. 
For this, the autocorrelation coefficient of streamwise velocity, $cc_u$, is calculated as:
\begin{equation} 
cc_u(x,y,z,\Delta x)=\frac{\overline{u'(x,y,z,t)u'(x+\Delta x,y,z,t)}}{\left(\overline{u'^2(x,y,z,t)}\right)^{0.5} \left(\overline{u'^2(x+\Delta,y,z,t)}^{0.5}\right)}.    
\label{coreq}
\end{equation} 
$cc_u$ is first obtained on the mid-span plane at various wall-normal locations within the boundary layer at $x=2, 2.25$ and $x=2.5$, from which, the integral length-scales at each streamwise location are calculated. 
The locus of GIP in $2 \le x \le 2.5$ (marked as $y_{GIP}$) can be approximated by the dashed horizontal line. 
At $x=2$, the near-wall region is dominated by fundamental and superharmonics in the 2D instability, thus having a relatively small length-scale. 
Below the GIP, the lambda vortices result in a local maxima in $L_{cx}$, at $y \sim 0.012$. 
Here, the lambda vortices are detached from the ``rollers'' on the wall, and the ``legs'' are inclined with respect to the GIP. 
Near the GIP and above it, $L_{cx}$ reduces because of the absence of any significant streamwise oriented structures. 
This trend is further enhanced at $x=2.25$ due to the stretching of the lambda vortices, which increases the local maxima in $L_{cx}$. 
As the lambda vortices penetrate the boundary layer, $L_{cx}$ exhibits a smoother profile as seen at $x=2.5$.
   
\section{Spectral-domain analysis of transition}\label{sec_spdmch}

\begin{figure}
\centering
\setlength\fboxsep{0pt}
\setlength\fboxrule{0pt}
\fbox{\includegraphics[width=6.0in]{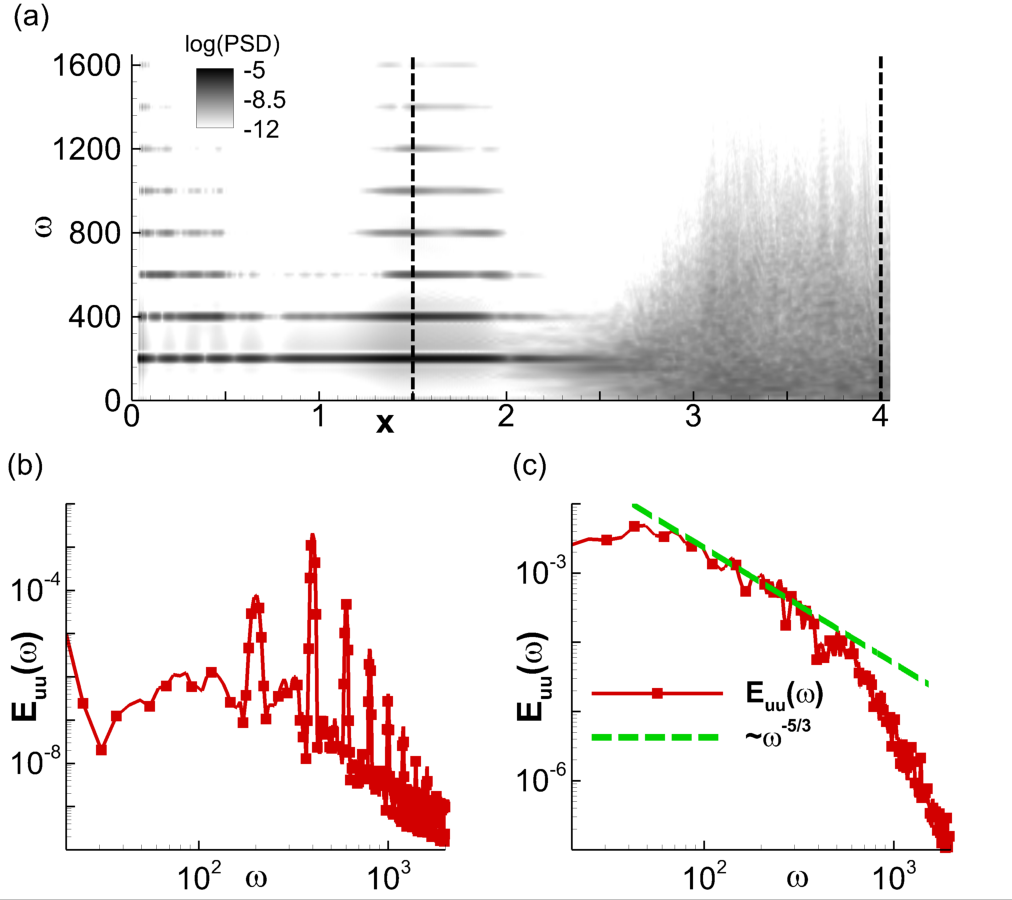}}
\caption{(a) Wall-pressure frequency content variation in terms of power spectral density obtained at various streamwise locations at the mid-span. 
Streamwise velocity spectra at (b) $x=1.5$ and $x=4$. 
Dashed line in (c) represents the slope, $\omega^{-5/3}$.}
\label{figftspc}
\end{figure}

The frequency spectrum of wall-pressure fluctuations provide a quantitative representation of the nonlinearity and transitional characteristics of the HBL. 
This is obtained on the mid-span and plotted in terms of the logarithm of power spectral density (PSD) in figure~\ref{figftspc}(a).
The horizontal axis is $x$ and the vertical axis represents circular frequency, $\omega$. 
The two vertical dotted lines mark $x=1.5$ and $x=4$. 
The spectrum near the leading edge shows the imprint of the actuator, primarily at $\omega =200$, from which multiple superharmonics develop due to nonlinear effects. 
As the BL grows, these superharmonics dampen, as seen in $0.5 \le x \le 1$. 
Once the linear instability region of the fundamental wave begins at $x \sim 1$ (figure~\ref{figeigsfs}(c)), the superharmonics also amplify ( $1 \le x \le 2$). 
At $x \sim 1.5$, most of the superharmonics exhibit peak amplitudes, and the streamwise velocity spectra, $E_{uu}(\omega)$, obtained here is plotted in figure~\ref{figftspc}(b). 
Nonlinear saturation limits the linear amplification of the fundamental, and peak energy is observed in the second superharmonic, $\omega =400$. 
Although the spectrum is narrow-banded (at integer multiples of the fundametal) at higher frequencies, the lower range shows a broadband nature, indicating percolation of frequencies to either side of the fundamental. 
This will be further examined in the context of nonlinear coupling of frequencies, below.

For $x>2$, the superharmonics above $\omega =400$ rapidly attenuates outside the region of linear instability. 
In $2 \le x \le 3$ there is a qualitative shift in the boundary layer character, where the harmonic narrow-band spectrum gives way to a broadband spectrum. 
As seen in the iso-level plots (figure~\ref{figqiso}), this region is characterized by oblique waves that distorts the 2D ``rollers'', and results in early turbulence with hairpin structures. 
Since $\omega =200$ is still the dominant frequency in $2 \le x \le 2.5$, it indicates that fundamental resonance \citep{sv2015direct} is the most probable cause of transition in this scenario. 
There are also traces of  peaks at $\omega \sim 150$ near the wall and $\omega \sim 100$ (near the GIP, not evident in this plot). 
The latter will be revisited in a following discussion on bicoherence, although the relatively lower energy content at the half-frequency of the fundamental significantly limits the effects of subharmonic resonance \citep{sv2014numerical}. 
This shows that fundamental resonance is the preferable mode of breakdown for a given 2D second-mode wave, in the presence of unbiased spanwise variations upstream. 
Fundamental resonance has also been observed to be naturally arising in HBLs over flared cones by \citet{ha2019dnsfc,ha2020threed}. 

At further downstream locations, $x>3$, the frequency spectrum is broadband, with no trace of dominance of the forcing-harmonic, or its multiples. 
The spectral slope is evaluated at $x=4$ using $E_{uu}(\omega)$, and is reported in figure~\ref{figftspc}(c). 
The dotted line marks the slope, $\omega^{-5/3}$, and shows that the HBL reaches a fully turbulent state towards the outflow boundary, with an inertial sub-range extending over a decade of frequencies. 

\begin{figure}
\centering
\setlength\fboxsep{0pt}
\setlength\fboxrule{0pt}
\fbox{\includegraphics[width=6.0in]{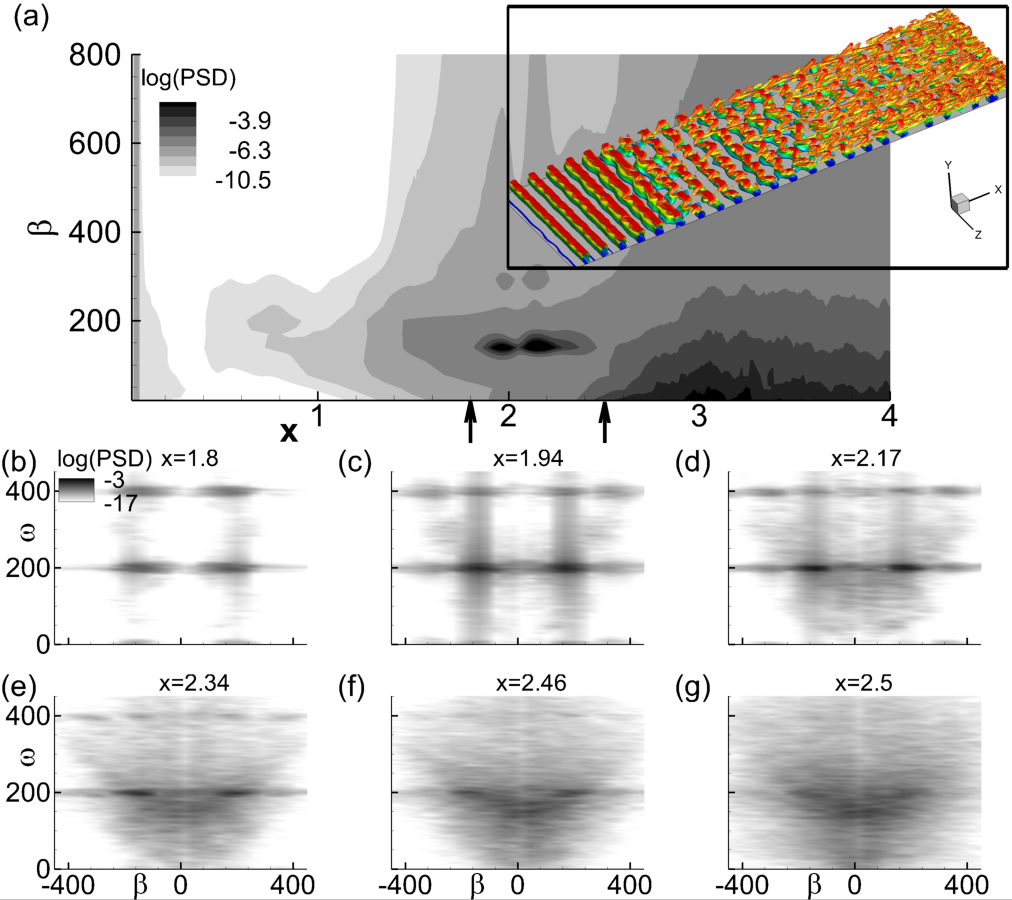}}
\caption{(a) Spanwise wavenumber spectra obtained from wall-pressure perturbations. 
Vertical arrows mark $1.8 \le x \le 2.5$. 
Inset shows Q-criterion colored with $u$ in $1.8 \le x \le 2.5$.
(b) - (g) Wavenumber-frequency spectra at indicated locations.}
\label{figftspcxk}
\end{figure}

The wavenumber spectra in figure~\ref{figftspcxk}(a) 
shows the sensitivity of the HBL to a narrow band of wavenumbers in the nonlinearly distorted upstream region, $x \le 1.5$. 
The horizontal axis is $x$ and the vertical axis represents spanwise wavenumber, $\beta$. 
These oblique modes ride on the ``rollers'' and modulate it in the spanwise direction, as the 2D wave amplifies. 
An interesting feature here is the region of peak amplification of the oblique mode, $1.8 \le x \le 2.5$, marked by two vertical arrows on the horizontal axis. 
The most amplified oblique mode has a spanwise wavenumber of $\beta \sim 150$, corresponding to a wavelength of $\lambda_z \sim 0.04$, which is $20\%$ of the span of the domain. 
For reference, the vortical structures in the HBL within this streamwise extent are also shown in an inset, using Q-criterion colored with $u$. 
This region of intense amplification of the oblique modes is crucial to the final breakdown of the boundary layer, through the formation of initial hairpin vortices. 
The significance of this streamwise region is limited in the linear framework, since the 2D wave monotonically attenuates beyond $x \sim 1.7$ (figure~\ref{fignnlndns}(d)). 
But the nonlinear response of the 2D wave indicates that there is a second region of amplification here (figure~\ref{fignnlndns}(d)), which is also evident in the fundamental 2D mode extracted through POD (figure~\ref{fignlpod}(a)). 
This region exhibits strong non-parallel, localized amplification of the fundamental frequency, which clearly harbors the oblique modes and makes the ``rollers'' susceptible to spanwise breakdown. 
There is a qualitative shift in the HBL in the vicinity of $x \sim 2.5$, as it remains no longer narrow-band in the wavenumber space. 
The wavenumber spectrum is rapidly populated near the lower end, as the HBL becomes turbulent towards the outflow.

Due to its key role in breakdown, the spectra in $1.8 \le x \le 2.5$ are further analyzed in the wavenumber-frequency ($\beta - \omega$) plane, at $6$ equally spaced streamwise locations, as shown in figures~\ref{figftspcxk}(b)-(g). 
The logarithm of PSD is plotted  to identify the spatio-temporal characteristics of the instabilities. 
The spanwise homogeneous component is removed prior to obtaining the transformation, to highlight the dynamics in the oblique modes. 
At $x=1.8$, the spectrum is still narrow banded in the frequency domain, with the fundamental and its superharmonic evident. 
At both these frequencies, the spatial scales are localized at $\beta \sim 150$. 
Progressing downstream, the superharmonic weakens, and the oblique mode at the fundamental frequency gains prominence by $x=1.94$ and $x=2.17$, which is another indication that the transition is dominated by the fundamental breakdown mechanism. 
In the latter half of the domain $1.8 \le x \le 2.5$ (figures~\ref{figftspcxk}(e), (f) and (g)), the spectrum rapidly broadens in the wavenumber-frequency domain. 
Energy eventually percolates to lower wavenumbers and frequencies due to the development of spanwise coherence in the near-wall structures and  initiation of quasi-streamwise vortices.

\begin{figure}
\centering
\setlength\fboxsep{0pt}
\setlength\fboxrule{0pt}
\fbox{\includegraphics[width=6.0in]{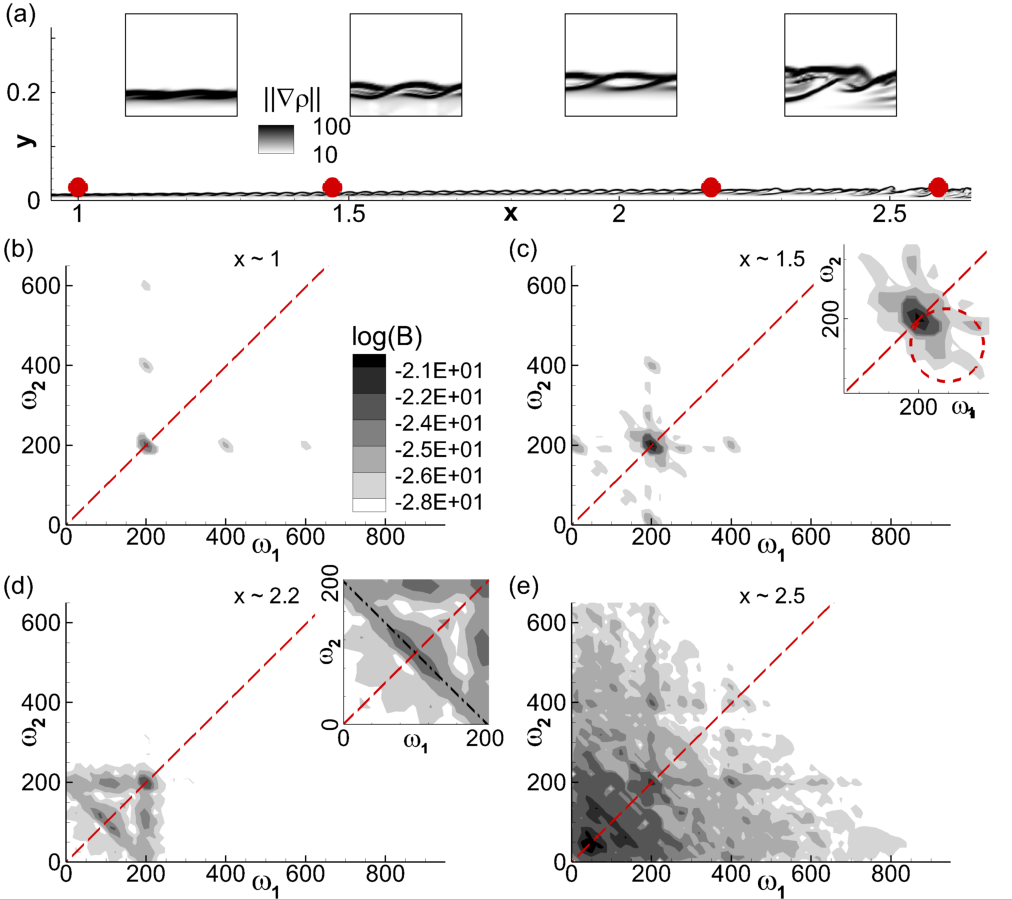}}
\caption{(a) Density-gradient magnitude contours, marked with the locations, $x \sim 1$, $x \sim 1.5$, $x \sim 2.2$ and $x \sim 2.6$. 
The vicinity of these locations are magnified in the $4$ insets. 
(b) - (f) Bispectrum evaluated at the above locations. 
The dotted inclined lines bisect the first quadrant. 
The dash-dot line in (d) represents $\omega_1 + \omega_2 = \omega_A$.}
\label{figbptrm}
\end{figure}

The above spectral analyses provide quantitative insights into the frequencies and wavenumbers present in the HBL. 
In order to understand the genesis of these frequencies, we need to identify those waves that are nonlinearly coupled, and the extent of their coupling (which results in spectral broadening), using higher order spectral quantities, like bispectrum. 
Bispectrum, $B(\omega_1,\omega_2)$, of a signal, $\phi(t)$, is defined as:
\begin{equation} 
B(\omega_1,\omega_2)=\lim_{T\to\infty}\frac{1}{T}E[\Phi(\omega_1)\Phi(\omega_2)\Phi^c(\omega_1+\omega_2)].
\label{coreq}
\end{equation} 
$T$ is the temporal duration of the signal, $\phi(t)$, $E[.]$ is the expectation operator, $\Phi(\omega)$ is Fourier transform of $\phi(t)$ and the superscript, $(.)^c$, represents the conjugate transpose. 
Below, we utilize bispectrum of pressure signals in the vicinity of the GIP, to track the nonlinear coupling leading to generation of new frequencies in the nonlinearly saturated and transitional regimes of the HBL. 
The results are presented in figure~\ref{figbptrm}. 
The density-gradient magnitude is plotted in figure~\ref{figbptrm}(a) to visualize the braided ``rope-like'' structures near the GIP, and eventual breakdown, downstream. 
The solid circles mark four locations where the time-traces of pressure are acquired to calculate the bispectrum. 
The state of the HBL in the vicinity of these locations are also  magnified in $4$ insets, for reference.
The braids are not evident at $x \sim 1$ since the nonlinearities are relatively weaker here. 
As the superharmonics amplify, the interlacing is more prominent at $x \sim 1.5$ and $x \sim 2.2$, consistent with the 2D nonlinear DNS. 
At $x \sim 2.6$, the GIP is no longer compact, and indicates nonlinear breakdown. 

Bispectrum at $x \sim 1$ is presented in figure~\ref{figbptrm}(b). 
The axes represent circular frequency, and the inclined dashed line marks the $45^o$ angle, bisecting this quadrant.
A support at $(\omega_1,\omega_2)$ in this plane suggests quadratic phase coupling among the waves with frequencies,  $\omega_1, \omega_2$ and $\omega_3 = \omega_1 + \omega_2$. 
With additional contextual data from the Fourier spectrum, bispectrum yields information about the specific (sum or difference) interaction among these three frequencies \citep{bountin2008evolution}. 
Due to the symmetry of bispectrum, it is sufficient to examine just one half of this quadrant. 
The most prominent nonlinear interaction at this location occurs at the fundamental $(\omega_1 = \omega_2 =200)$, to produce its superharmonic ($\omega_3 = 400$). 
The fundamental also interacts with the superharmonics thus produced, {\em e.g.}, $(\omega_1 = 200, \omega_2 =400)$ and $(\omega_1 = 200, \omega_2 =600)$, resulting in the banded spectrum discussed earlier in figure~\ref{figftspc}(a).
Similar nonlinear interactions of second-mode have been reported in bispectrum analysis of experimental  data in HBLs by \citet{kimmel1991nonlinear,chokani1999nonlinear}. 
This resulted in a nonlinear regime dominated by fundamental resonance, which is also the case here. 

The bispectrum at $x \sim 1.5$ (figure~\ref{figbptrm}(c)) is not as compact as at the previous location. 
The higher superharmonics are no longer significant here, with the coupling mainly limited to self interaction of the fundamental and $(\omega_1 = 200, \omega_2 =400)$. 
An interesting feature here is the spectral support in the vicinity of the self interaction. 
This region is marked by a dashed circle in the inset in figure~\ref{figbptrm}(c). 
It mostly corresponds to interactions of the type, $(\omega_1 = 200+\Delta, \omega_2 =200-\Delta)$, resulting in the superharmonic, $\omega_3 = 400$. 
$\Delta$ is a small deviation from the harmonic. 
This was observed in the nonlinear regime of second-mode instability over a flared cone in the experiments of \citet{craig2019nonlinear}.
Along the $45^o$ line, the interactions favor $(\omega_1 = 200+\Delta, \omega_2 =200 + \Delta)$, resulting in $\omega_3 = 400 + 2\Delta$, representing the ``broadband self interaction'' \citep{craig2019nonlinear} of second-mode instability, resulting in spectral broadening. 
By this streamwise location we also notice interaction of the fundamental with the neighborhood of the zero frequency. 
This is interpreted as the nonlinear interaction leading to energy transfer from the mean flow into the fundamental frequency. 
This has also been found \citep{craik1971non} to result in strong amplification of the oblique wave, thus inducing spanwise periodicity. 
In figure~\ref{figftspcxk}(a), we indeed observe that the nonlinearly saturated second-mode increasingly become selective to oblique mode instabilities in this streamwise region, converging on to $\lambda_z \sim 0.04$ as the dominant wavelength prior to breakdown. 
There is also a possibility of this interaction to induce gradual modulation in the second-mode envelope as discussed in \citet{craig2019nonlinear}, which is also the case here.

At $x \sim 2.2$ (figure~\ref{figbptrm}(d)) , another interesting quadratic coupling is evident in the signal. 
There is a continuous range of frequencies involved in the interaction, as defined by the linear equation, $\omega_1 + \omega_2 = \omega_A$, with $\omega_A$ being the actuator frequency as defined earlier. 
This line segment is marked in the magnified image (inset in figure~\ref{figbptrm}(d)). 
The highest degree of coupling is observed between the frequencies,  $(\omega_1 = 90, \omega_2 =110)$, which are close to the subharmonic of the fundamental frequency. 
Such a continuous linear range of interaction and the small offset of peak interaction region (from the subharmonic) has been reported by \citet{bountin2008evolution}. 
The coupling of the second-mode fundamental with its fist subharmonic is essential for the subharmonic-resonance route to transition (see {\em} \citet{shiplyuk2003nonlinear}). 
Thus although fundametal resonance is observed to be the primary mechanism of transition through spectral analysis, the quadratic phase coupling identifies the presence of subharmonic resonance as well, in the later nonlinear stages. 
This is possible because of the unbiased forcing of spanwise variations in the 2D second-mode, providing necessary scales for both the routes.
The interaction of the fundamental with the subharmonic, $(\omega_1 = 200, \omega_2 =100)$, generating $\omega_3 = 300$ is evident in the $x=St$ plot (figure~\ref{figftspc}(a)) as a weak spectral peak in the streamwise extent, $2 \le x \le 2.4$.
Following the formation of the interaction, $\omega_1 + \omega_2 = \omega_A$, the bispectrum is quickly populated towards the lower frequencies, as seen at 
$x \sim 2.6$ (figure~\ref{figbptrm}(e)). 
By this location the Fourier spectrum also attains a broad peak at frequencies below the fundamental, and the boundary layer is in the later stages of transition. 
Here, cubic and higher order interactions could become relevant among the wide range of frequencies present. 

\begin{figure}
\centering
\setlength\fboxsep{0pt}
\setlength\fboxrule{0pt}
\fbox{\includegraphics[width=6.0in]{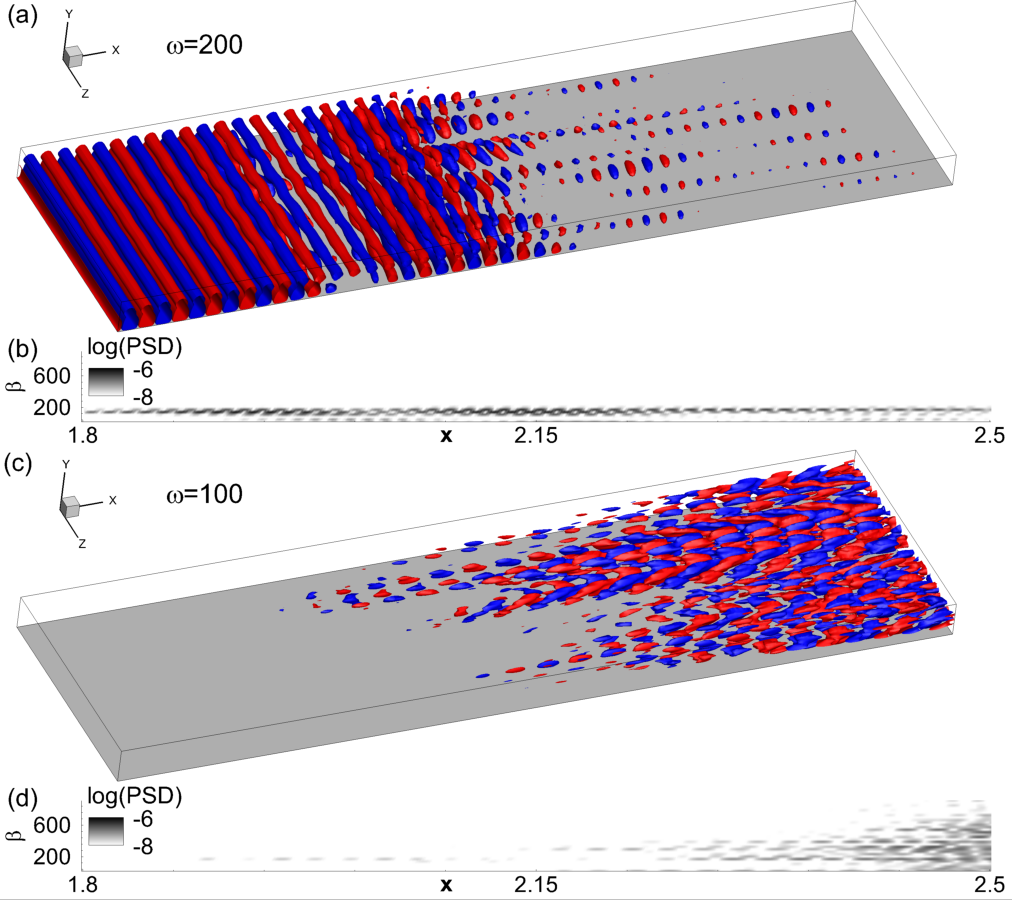}}
\caption{Iso-levels of $u'$ used to represent the (a) fundamental and (c) subharmonic modes in the transitional region. 
Spanwise wavenumber spectrum of the (b) fundamental and (d) subharmonic modes in the transitional region.}
\label{fidvisodmd}
\end{figure}

Since the fundamental and its subharmonic are found to be relevant in the transitional region, we extract their global forms in figure~\ref{fidvisodmd}. 
The 3D structures of the pure-frequency modes are extracted using the dynamic mode decomposition \citep{schmid2010dynamic} (DMD), which also approximates the Koopman modes of the nonlinear operator \citep{rowley2009spectral}. 
Thus, although this flowfield is nonlinearly saturated, these modes represent the relevant eigenfunctions of the infinite-dimensional linear operator that approximates the nonlinear evolution of the flow. 
The iso-levels of the fundamental frequency in $u'$ is shown in figure~\ref{fidvisodmd}(a). 
Its spanwise wavenmumber spectrum (with the $\beta=0$ component removed) obtained at the wall is provided in figure~\ref{fidvisodmd}(b) for reference. 
The axial extent, $1.8 \le x \le 2.5$, is similar to that discussed in figure~\ref{figftspcxk}. 
The upstream 2D wave gradually develops oblique waves and disintegrate in the spanwise direction. 
The spatial support of the fundamental mode attenuates significantly beyond $x \sim 2.15$. 
The spanwise wavenumber has a narrow spectral range and corresponds to its peak value observed earlier in figure~\ref{figftspcxk}(a). 
This confirms that $\lambda_z \sim 0.04$ is the dominant instability of the second-mode ``rollers'' at the fundamental frequency. 
The subharmonic mode and its spanwise spectrum are presented in figures~\ref{fidvisodmd}(c) and (d), respectively. 
It becomes significant at $x > 2.15$, consistent with the quadratic interaction detected in figure~\ref{figbptrm}(d). 
This instability begins at a spanwise wavenumber identical to that in the fundamental, but quickly populates the spanwise spectrum towards the breakdown region. 
In the following section we study the post-breakdown regime to explore the effects on the wall.

\section{Turbulent HBL and near-wall effects}\label{sec_spdmch}

\begin{figure}
\centering
\setlength\fboxsep{0pt}
\setlength\fboxrule{0pt}
\fbox{\includegraphics[width=6.0in]{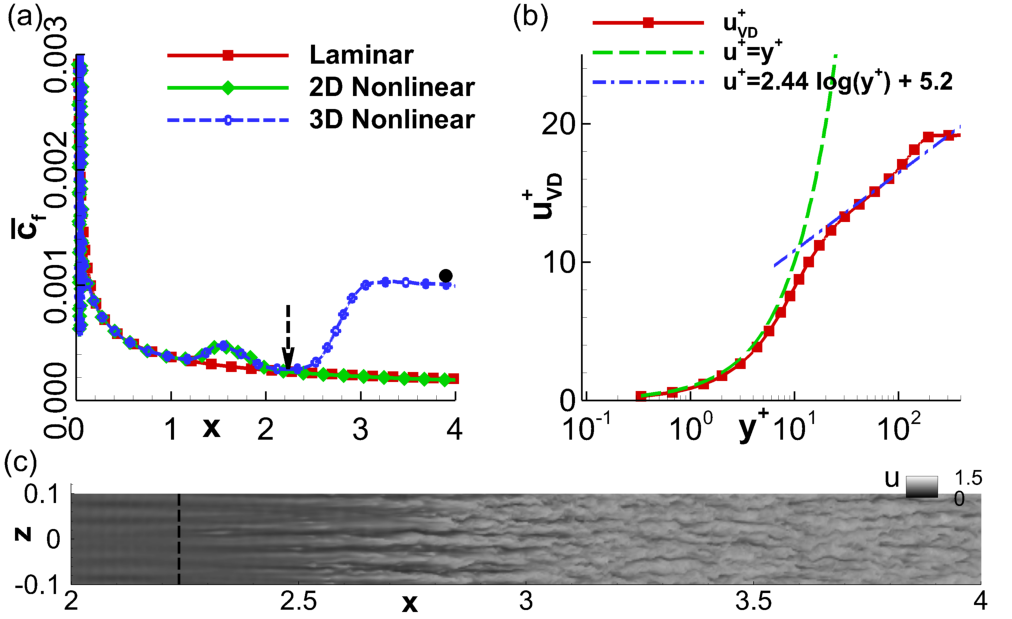}}
\caption{(a) Time-averaged skin friction coefficient for various cases as indicated. 
(b) Wall-normal velocity profile at $x=4$, in terms of wall units and friction velocity. 
(c) Contours of $u$ on a wall-parallel plane at $y \sim 8 \times 10^{-3}$.}
\label{figlwal}
\end{figure}

The impact of the state of HBL on the wall is readily identifiable through the skin friction coefficient, $c_f$, defined as: 
\begin{equation}\label{cfeqs}
c_f=\frac{2}{Re} \mu \left. \frac{\partial u}{\partial y} \right |_{y=0}.
\end{equation}
The time-averaged skin friction coefficient, $\overline{c}_f$, is plotted in figure~\ref{figlwal}(a), 
for the 2D laminar, 2D nonlinear and 3D nonlinear simulations, as indicated. 
For the 3D case, its value is also averaged across the span. 
The laminar value of $\overline{c}_f$ monotonically decreases in the downstream direction, following its peak induced due to the viscous-inviscid interaction near the leading edge shock wave. 
The $\overline{c}_f$ obtained for the nonlinearly saturated 2D DNS is very similar to that of the laminar case, except for a localized hump in $1 \le x \le 2$ corresponding to the peak amplitude of the saturated second-mode waves. 
Such a peak in skin friction was also observed by \citet{franko2013breakdown} in the region of second-mode saturation. 
The 3D DNS closely follows the trend in the 2D nonlinear DNS until this hump relaxes to the laminar value. 
After this location, the skin friction rapidly rises due to the spanwise breakdown of the boundary layer, leading to transition. 
The location of this local minima in  $\overline{c}_f$ is obtained as $x \sim 2.24$ (marked with a dashed arrow), and can be considered \citep{kimmel1993effect} as an indicator of the location at which transition begins. 
This location also matches with the streamwise region beyond which the spanwise spectra switches-over from a narrowband form (peak energy near $\beta \sim 150$ or $\lambda_z \sim 0.04$) to one where a wide range of low wavenumbers are energized (discussed previously in figure~\ref{figftspcxk}(a)). 
By the end of the domain, $x=4$, the coefficient is $\sim 0.001$, which is consistent with the turbulent skin friction value reported in \citet{egorov2016direct} for identical freestream conditions (marked with a solid circle). 
The wall-normal profile of streamwise velocity in terms of wall units ($y^+$) and friction velocity ($u^+$) at this location is provided in figure~\ref{figlwal}(b). 
The van Driest transformation is applied to obtain $u^{+}_{VD}$, which encompasses the enhanced effects of compressible fluctuations at this edge Mach number\citep{von1956problem,schetz2011boundary}. 
The dashed curve is a linear relation expected in the viscous sublayer, which extends till around $y^+ \sim 10$, consistent with prior calculations. \cite{franko2013breakdown,sv2015direct}. 
The viscous sublayer is followed by the buffer layer, beyond which, a logarithmic variation is observed in the velocity profile. 
This profile is compared to a reference log-law formula \citep{roy2006review}, which indicates the existence of a turbulent boundary layer. 
The velocity defect law constitutes the deviation from the log law in the outer region.  

In cases where second-mode fundamental resonance dominates, the transition location is relatively prolonged \citep{koevary2010numerical,khotyanovsky2016numerical}, which has been associated to the weaker streamwise streaks generated \citep{franko2013breakdown}. 
These streaks are visible in figure~\ref{figlwal}(c), primarily in $2 \le x \le 3$, which plots the instantaneous streamwise velocity contours at $y \sim 8 \times 10^{-3}$. 
This wall-normal location corresponds to $y^+ \sim 30$ (lower end of the log layer), based on the boundary layer properties at the exit of the domain ($x=4$). 
The transition location, $x = 2.24$, is also marked with a dashed vertical line. 
Consistent with the results in \citet{franko2013breakdown}, the streamwise streaks are generated following the saturation of the fundamental second-mode, and eventually develop undulations (observed within $2.5 \le x \le 3$), before disintegrating in the turbulent region downstream. 
Although the fully turbulent region resulting from second-mode fundamental resonance was not addressed in the above reference, detailed analysis of the transitional regime indicated that no peak existed in the skin friction plot. 
Figure~\ref{figlwal}(c) also shows that the instantaneous undulations in the streaks originate at slightly different locations along the span of the plate. 
This could explain why skin friction overshoot was not observed in the time- and span-averaged $c_f$, plotted in figure~\ref{figlwal}(a). 
A previous study \citep{unnikrishnan2019first} of this HBL where transition was induced through forcing by a monochromatic first-mode oblique wave, clearly identified a skin friction overshoot in the transitional regime. 

\begin{figure}
\centering
\setlength\fboxsep{0pt}
\setlength\fboxrule{0pt}
\fbox{\includegraphics[width=6.0in]{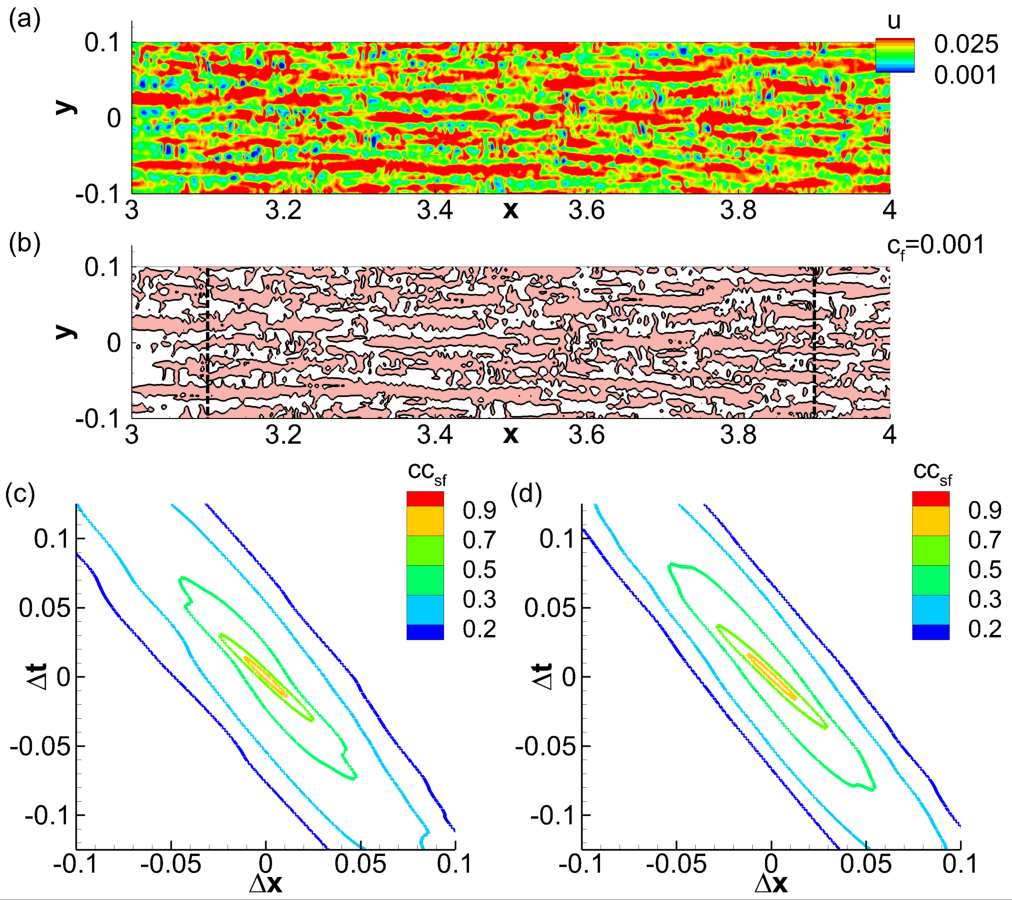}}
\caption{(a) Instantaneous contours of $u$ near the wall. 
(b) Corresponding signature of $c_f$, plotted using the contour representing its mean value in the turbulent region. 
Spatio-temporal correlations in $c_f$ obtained at (c) $x=3.1$ and (d) $x=3.9$.}
\label{figuccfc}
\end{figure}

Near-wall features in the boundary layer provide significant insights into the drag penalty on the surface. 
Here, we study the influence of the transitioned HBL on the wall, as it evolves into a turbulent state in  $3 \le x \le 4$. 
The streamwise velocity distribution near the wall is a relevant quantity to identify the skin friction characteristics, and are presented in figure~\ref{figuccfc}(a).
The instantaneous contours indicate the presence of several near-wall patches of high values of $u$. 
Being a surrogate for $c_f$, this qualitatively indicates regions of high skin-friction. 
To related these regions to shear events, the corresponding instantaneous $c_f$ contours are plotted in  figure~\ref{figuccfc}(b). 
The contour level chosen here is $c_f=0.001$ which is approximately the value of the mean skin friction eventually attained in the turbulent region, as seen in figure~\ref{figlwal}(a).
The large-scale patterns in near-wall velocity contours have corresponding signatures in the above-average skin friction regions, consistent with the observations of \citet{pan2018extremely}, where regions of extreme skin friction  were characterized by large-scale structures. 
Their conditional analysis identified ``finger-shaped'' large-scale structures in negative wall-normal velocity perturbations directly above these high-shear regions, which signify ``splatting'' events \citep{agostini2016skewness}, constituting strong movements of outer-layer high-momentum fluid towards the wall.

The post-transitional characteristics of the HBL is expected to evolve until fully turbulent conditions are achieved. 
The evolution of skin-friction patterns in this region are further quantified using spatio-temporal correlations, $cc_{sf}$ in figure~\ref{figuccfc}(c) and (d). 
They are obtained as:
\begin{equation} 
cc_{sf}(x,y,z,\Delta x,\Delta t)=\frac{\overline{c_f(x,y,z,t)c_f(x+\Delta x,y,z,t+\Delta t)}}{\left(\overline{c_f^2(x,y,z,t)}\right)^{0.5} \left(\overline{c_f^2(x+\Delta,y,z,t+\Delta t)}^{0.5}\right)}.    
\label{coreq}
\end{equation}
The spanwise-averaged results correspond to $x=3.1$ and $x=3.9$, which highlight key changes in skin friction patterns that have transpired within this streamwise extent. 
The primary observation is that the width of the spatio-temporal correlation contours decreases as the HBL becomes turbulent. 
Here, the decrease is about $10\%$, between the two locations examined. 
This indicates a reduction in the length of the high-velocity patches near the surface. 
No significant changes were observed in the convection velocity of these patches, as suggested by the similar slopes of the correlation contours at these two locations. 

\begin{figure}
\centering
\setlength\fboxsep{0pt}
\setlength\fboxrule{0pt}
\fbox{\includegraphics[width=6.0in]{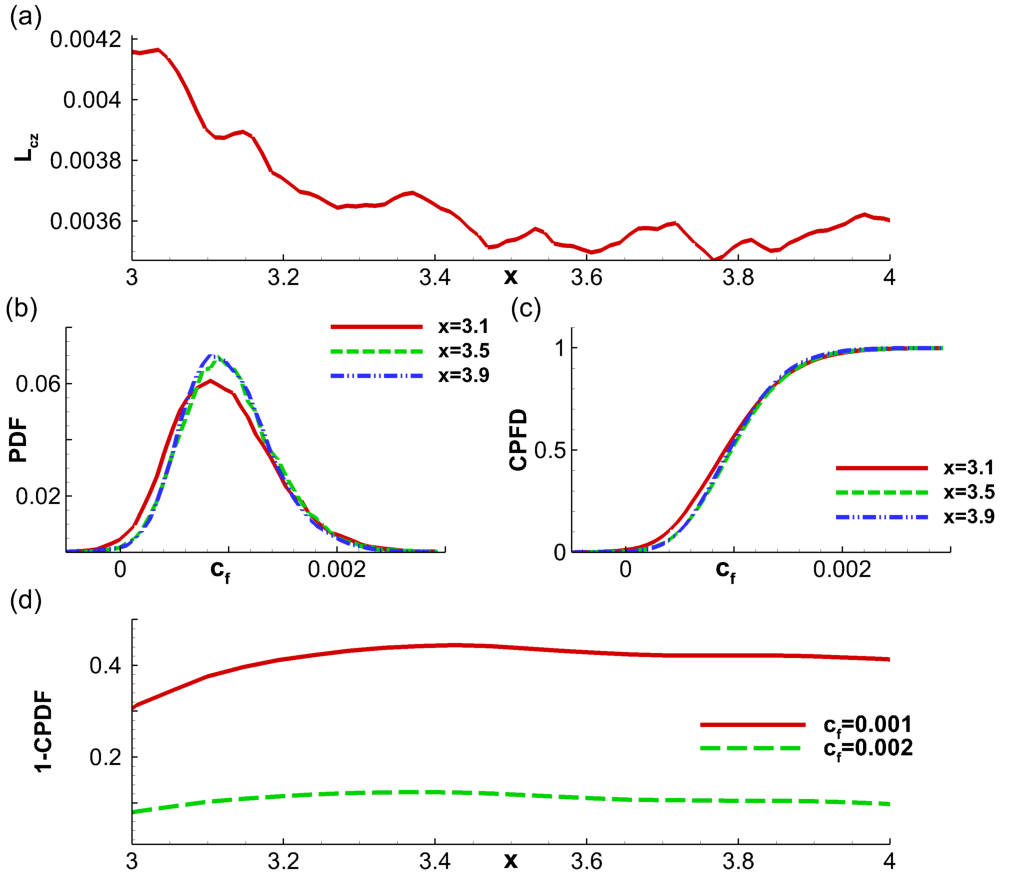}}
\caption{(a) Spanwise integral length-scale in $c_f$ plotted with $x$. 
(b) PDF of $c_f$ at indicated locations. 
(c) Corresponding CPFs, as indicated. 
(d) Probability of occurrence of shear events above the threshold value (as indicated) plotted with $x$.}
\label{figcrpdf}
\end{figure}

The spanwise extent of the regions of high skin friction also varies in the boundary layer, as indicated by the integral length-scale, $L_{cz}$, in figure~\ref{figcrpdf}(a),  
plotted as a function of streamwise distance. 
$L_{cz}$ is obtained by integrating the autocorrelation function of $c_f$. 
A decrease in spanwise width of around $14 \%$ is observed within  $3 \le x \le 3.6$, beyond which fluctuations induce only minor variations. 
The high skin friction patches initially display spanwise extents that closely follow the secondary instability of the ``roller'' structures, which were disintegrated following the amplification of oblique waves with $\lambda_z \sim 0.04$. 
 
The occurrence of shear events can also be quantified using their probability distribution function (PDF) at various streamwise locations. 
Spanwise averaged PDFs are provided in figure~\ref{figcrpdf}(b), at the indicated locations. 
Immediately following transition, $c_f$ exhibits a relatively broader distribution, at $x=3.1$, with a positive skewness (mean $\sim 0.001$, mode $\sim 0.0008$). 
As turbulence sets in, the $c_f$ distribution becomes narrower, with the PDFs at $x=3.5$ and $x=3.9$ exhibiting greater occurrences of skin friction patches around the mean value of $c_F \sim 0.001$. 
The cumulative PDFs (CPDF) of $c_f$ at these three locations (figure~\ref{figcrpdf}(c)) indicate that developing turbulence in the boundary layer results in a reduction of events with low skin friction impact, in $0 \le c_f \le 0.001$. 
These plots also confirm that the distribution characteristics of wall loading have converged to an equilibrium state in the latter half of the domain, $3 \le x \le 4$. 
Since the positive tail in the streamwise-velocity and $c_f$ distribution is indicative of large-scale sweeping motions \citep{agostini2016skewness}, the CPDF is used to quantify the occurrence of such events in the post-transition region. 
The quantity, $1-CPDF$, at a chosen value of $c_f$, shows the fractional occurrence of events above that threshold. 
Figure~\ref{figcrpdf}(d) plots this function with x for two values of $c_f$, one near the mean and the other twice that value. 
The plot for $c_f=0.001$ indicates that large-scale near-wall sweeping motions increase following transition, and eventually equilibrate. 
The occurance of very extreme skin friction events categorized as above $c_f=0.002$ exhibit relatively lesser variations, but nevertheless settles to around $10\%$ by the end of the domain. 
This is consistent with the classification of large-scale events by \citet{agostini2014influence}, where a $10\%$ threshold was chosen for this segregation.

\section{Summary}\label{sec_cln}
The linear and nonlinear development of a 2D second-mode instability, followed by its role in transition and eventual generation of turbulence in a hypersonic boundary layer (HBL) is studied using linear stability theory (LST) and direct numerical simulations (DNSs). 
FS~synchronization identifies the unstable mode~S which manifests as the amplified second-mode in the Mach~$6$ flat-plate boundary layer under consideration. 
This wave is recreated in a linearly forced 2D DNS, and a good match with LST predictions is observed, and serves as a benchmark for subsequent nonlinear DNSs. 
The linear density-perturbation field identifies ``rope-shaped'' structures along the generalized inflection point (GIP) of the base flow, consistent with experimental observations and prior computations. 
Upon imposing nonlinear forcing, the second-mode envelope becomes asymmetrically modulated and develop wider crests and narrower peaks in pressure perturbations, due to the generation of superharmonics and base-flow distortion. 
The density-perturbation field now exhibits tightly braided structures better resembling schlieren visualizations of late-stage evolution of second-mode wavepackets. 
The saturated field is delineated into the orthogonal modes, which identify a second region of growth of the fundamental frequency, outside the linear envelope.  
The integer superharmonics thus extracted, exhibit a proportional decrease in wavelengths, thus maintain similar phase-speeds as the fundamental mode. 

The 3D DNS forces the 2D second-mode in the presence of weak random perturbations, thus allowing the receptivity of the nonlinearly distorted base flow to guide the breakdown process. 
The secondary instability of the 2D ``roller'' structures causes selective amplification of a narrow band of spanwise wavenumbers, resulting in lambda vortices below the GIP and spanwise disintegration. 
These lambda vortices are detached from the ``rollers'' beneath it, which exhibit relatively weaker spanwise modulations. 
The streamwise vortex-stretching in lambda vortices generates hairpin vortices, at an ``inclination'' to the GIP, resulting in a localized peak (below the GIP) in streamwise length-scales within the transitional HBL. 

Frequency spectrum shows localized generation of superharmonics within the region of linear instability, outside of which, it attenuate. 
Post spanwise-breakdown, subharmonic and lower frequencies are excited, and a broadband nature is observed. 
Towards the end of the domain, the streamwise-velocity spectrum exhibits a well-developed inertial sub-range, suggesting a fully turbulent boundary layer.
The spanwise wavenumber spectrum identifies the sensitivity of the nonlinearly saturated HBL to oblique waves with wavelengths around $20\%$ of the spanwise extent of the domain. 
The peak energy in this wave is attained in the second region of amplification of the fundamental frequency identified in the 2D nonlinear simulation, and results in spanwise breakdown, quickly populating the wavenumber spectrum. 
Presence of fundamental resonance is confirmed through the observation of these oblique waves, which occur at the fundamental frequency. 
Analysis of the pressure signal near the GIP through bispectrum yields key insights into the quadratic coupling present in the nonlinear flowfield, consistent with prior experiments. 
The fundamental progressively interacts with itself and the superharmonics thus generated, resulting in a multi-harmonic spectrum prior to transition. 
This is followed by ``broadband self interaction'' at frequencies in the vicinity of the fundamental. 
The experimentally observed interactions among a continuous range frequencies defined by a linear equation are also identified further downstream, which also detects the presence of subharmonic resonance in the HBL. 
Modal analysis yields the global form of the fundamental and subharmonic components - the former constitutes the most dominant feature until spanwise breakdown is initiated. 
The subharmonic indicates minimal presence prior to this, but emerges as the dominant component immediately downstream. 

Near-wall analysis of post-transition region shows that the skin friction coefficient, $c_f$, reaches the turbulent value for this freestream condition towards the end of the domain, where a turbulent $u$-velocity profile is observed.  
No localized peak is observed in the time-averaged $c_f$ in the transitional region, which is associated to the slight variations in the location where streamwise streaks experience breakdown. 
Regions of intense skin friction are identified, which correspond to the large-scale ``splatting'' motions entraining high-momentum fluid towards the wall. 
Spanwise and streamwise extents of these regions decrease following transition, until turbulence establishes in the boundary layer.
Thus fundamental resonance is found to result in a fully turbulent HBL downstream.

\section*{Acknowledgments}
This research was supported by the Office of Naval Research (Grants: N0001-13-1-0534, N00014-17-1-2584) monitored by E. Marineau, with R. Burnes serving as technical point of contact.
The simulations were performed with a grant of computer time from the DoD HPCMP DSRCs at AFRL, NAVO and ERDC, and the Ohio Supercomputer Center.

\section*{References}
\bibliography{MASTER_SECONDMODE}

\end{document}